\documentclass[seceq]{ptptex}
\usepackage{graphicx}
\usepackage{bm}

\newcommand{\Eq}[1]{Eq.~({\protect\ref{#1}})}

\newcommand{\Fig}[1]{Fig.~\protect\ref{#1}}

\newlength{\Tatescale}
\newlength{\figwidth}
\newcommand{\Cut}[1]{}
 
\newcommand{\beq}{\begin{eqnarray}}
\newcommand{\eeq}{\end{eqnarray}}

\newcommand{\bn}{\mbox{\boldmath $n$}}

\newcommand{\bq}{\mbox{\boldmath $q$}}
\newcommand{\bp}{\mbox{\boldmath $p$}}
\newcommand{\bk}{\mbox{\boldmath $k$}}
\newcommand{\br}{\mbox{\boldmath $r$}}
\newcommand{\bx}{\mbox{\boldmath $x$}}
\newcommand{\by}{\mbox{\boldmath $y$}}

\newcommand{\bv}{\mbox{\boldmath $v$}}
\newcommand{\bL}{\mbox{\boldmath $L$}}
\newcommand{\bJ}{\mbox{\boldmath $J$}}
\newcommand{\bP}{\mbox{\boldmath $P$}}
\newcommand{\bQ}{\mbox{\boldmath $Q$}}
\newcommand{\bS}{\mbox{\boldmath $S$}}

\newcommand{\bsigma}{\mbox{\boldmath $\sigma$}}
\newcommand{\btau}{\mbox{\boldmath $\tau$}}
\newcommand{\brho}{\mbox{\boldmath $\rho$}}

\newcommand{\bns}{\mbox{\scriptsize \boldmath $n$}}
\newcommand{\brs}{\mbox{\scriptsize \boldmath $r$}}
\newcommand{\bxs}{\mbox{\scriptsize \boldmath $x$}}

\newcommand{\bps}{\mbox{\scriptsize \boldmath $p$}}
\newcommand{\bqs}{\mbox{\scriptsize \boldmath $q$}}
\newcommand{\bks}{\mbox{\scriptsize \boldmath $k$}}

\newcommand{\pslash}{p\kern-1ex /}
\newcommand{\kslash}{k\kern-1ex /}
\newcommand{\qslash}{q\kern-1ex /}
\newcommand{\lslash}{l\kern-1ex /}
\newcommand{\sslash}{s\kern-1ex /}
\newcommand{\paslash}{p_a\kern-2ex /}
\newcommand{\pbslash}{p_b\kern-2ex /}
\newcommand{\Dslash}{{\cal D}\kern-1.5ex /}
\newcommand{\dslash}{\partial\kern-1.2ex /}

\pubinfo{Vol.~123, No.~1, January 2010}

\notypesetlogo 
\title{Theoretical Foundation of the Nuclear Force in QCD\\
and Its Applications to Central and Tensor Forces\\
in Quenched Lattice QCD Simulations}

\author{Sinya \textsc{Aoki},$^1$  Tetsuo \textsc{Hatsuda}$^2$ 
and Noriyoshi \textsc{Ishii}$^2$}

\inst{$^1$Graduate School of Pure and Applied Sciences,
University of Tsukuba, \\ Tsukuba 305-8571, Japan\\ 
$^2$Department of Physics, University of Tokyo, \\ 
Tokyo 113-0033, Japan}

\recdate{October 5, 2009}% 

\abst{We present full accounts of a method to extract 
nucleon-nucleon ($NN$) potentials from  the Bethe-Salpter amplitude  in
lattice QCD.  The method is applied to two nucleons on the lattice  
with quenched  QCD simulations. By disentangling the mixing between 
the S-state and the  D-state,  we obtain central and tensor potentials 
in the leading order  of the velocity expansion of the non-local $NN$ 
potential.
The  spatial structure and the quark mass dependence of the potentials 
are analyzed in detail.  
}

\PTPindex{164, 232, 234}

\begin{document}
\maketitle

%%%%%%%%%%%%%%%%%%%%%%%%%%%%%%%%%%%%%%%%%%%%%%%%%%%%%%%%%%%%%%%%%%%%%%%%%%%%%%%
\section{Introduction}
\label{sec:intro}
%%%%%%%%%%%%%%%%%%%%%%%%%%%%%%%%%%%%%%%%%%%%%%%%%%%%%%%%%%%%%%%%%%%%%%%%%%%%%%%

 The origin of the nuclear force is one of the 
 major unsolved problems in particle and nuclear physics
  even after the establishment of the quantum chromodynamics (QCD).
 Although the nuclear force is still not well-understood theoretically,
 a large number of proton-proton and neutron-proton
  scattering data as well as deuteron properties 
  have been  accumulated and
  summarized e.g. in the Nijmegen database \cite{Nijmegen_data}.
   To describe the  elastic nucleon-nucleon ($NN$) scattering at low-energies
 below the pion production threshold together with the deuteron properties,
 the notion of the $NN$ potential (either in the coordinate space or in
  the momentum space) turns out to be very useful \cite{NN-review}:
   it  can be determined phenomenologically to reproduce
  the scattering phase shifts and bound state properties 
  either through the Schr\"{o}dinger equation for the $NN$  wave function or through the 
  Lippmann-Schwinger equation for the $NN$  $T$-matrix.  Once the potential is determined,
   it can be
  used to study systems with more than 2 nucleons by using various many-body techniques.

 Phenomenological  $NN$  potentials which can fit
 the $NN$  data precisely (e.g. more than 2000 data points
  with $\chi^2/{\rm dof} \simeq 1$) 
 at $T_{\rm lab} < 300 $~MeV are called the 
 high-precision $NN$  potentials:
 They  include the potentials such as  CD-Bonn \cite{CD-Bonn},
   Argonne $v_{18}$ \cite{Wiringa:1994wb}, and 
  Nijm I, Nijm II and Reid93 \cite{Stoks:1994wp}.
 Also systematic low energy construction of the 
 nuclear force on the basis of the chiral perturbation theory 
 is being developed \cite{ChPT-1,ChPT-2}.

 The phenomenological  $NN$  potentials in the coordinate 
 space are known to reflect some characteristic features of the $NN$  interaction
 at different length scales \cite{NN-review}:
\begin{enumerate}
  \item[(i)]
  The long range part of the nuclear force  (the relative distance 
 $r >  2$  fm) 
 is dominated by the one-pion exchange introduced by Yukawa \cite{yukawa}.
  Because of the pion's Nambu-Goldstone character, it
  couples to the spin-isospin density of the nucleon and hence
  leads to a strong spin-isospin dependent force, namely the tensor force.
 \item[(ii)]  The medium range part ($1\ {\rm fm} < r < 2$ fm) receives 
  significant contributions from the exchange of  
  two-pions ($\pi\pi$) and heavy mesons ($\rho$, $\omega$, and $\sigma$).
  In particular, the spin-isospin independent attraction
 of about 50 -- 100 MeV in this region plays an essential role
  for the binding of atomic nuclei.
\item[(iii)]  The short  range part ($r < 1$ fm) is best described by
  a strong repulsive core as originally introduced by Jastrow \cite{jastrow}.
  Such a short range repulsion is important for  
  the stability of atomic nuclei against collapse, 
 for determining the maximum mass of neutron stars, and for
 igniting the Type II supernova explosions \cite{VJ}.
\item[(iv)]  There is also a strong attractive spin-orbit force in the 
  isospin 1 channel at medium and
 short distances. This leads to the $^3 {\rm P}_2$ neutron pairing
 in neutron matter and hence the neutron
  superfluidity inside neutron stars \cite{VJ}.
\end{enumerate}   
   
 A repulsive core  surrounded by an attractive well  is in fact a
  common feature of the ``effective" potential between
 composite particles.  The Lenard-Jones potential between 
  neutral atoms or molecules is a well-known example in 
  atomic physics. The potential between $^4$He nuclei 
  is a typical example in nuclear physics.
  The origin of the repulsive cores in these examples
 is  known to be the Pauli exclusion among electrons or among nucleons.
  The same  idea, however, is not directly applicable to the $NN$ potential,
  because the quark has not only spin and  flavor
  but also color which allows 
  six quarks to occupy the same state without violating  the Pauli principle. 
  To account for the repulsive core of the $NN$ force, therefore, 
  various ideas have been proposed as summarized in Ref.~\citen{QQ_review}:
   exchange of the neutral
  $\omega$ meson \cite{nambu57},  exchange of 
   non-linear pion field \cite{skyrme}, and a combination of the Pauli principle
    with the one-gluon-exchange between quarks  \cite{core_quark}.
  Despite all these efforts, convincing account of the 
  nuclear force has not yet been obtained.

  In this situation, it is highly desirable to
  study the $NN$ interactions from the first principle
  lattice QCD simulations. A theoretical framework
  suitable for such purpose was first proposed 
  by L\"{u}scher \cite{luescher}: For two hadrons in a finite
  box with the size $L \times L \times L$ in periodic boundary conditions,  
  an exact relation between  the energy spectra in the box
  and the elastic scattering phase shift at these energies was
   derived: If the range of the hadron interaction $R$  is sufficiently
  smaller than the size of the box $R<L/2$, the behavior of the 
  two-particle Bethe-Salpeter (BS)
  wave function $\psi (\br)$ in the interval $R < | \br | < L/2 $
  under the periodic boundary conditions
  has sufficient information to relate the phase shift and the 
  two-particle spectrum. 
  
  L\"{u}scher's method bypasses
  the difficulty to treat the real-time scattering process
  on the Euclidean 
  lattice.\footnote{
%
%%   If one  of the three quarks  inside the baryon  is infinitely heavy,
%% one  may define the  potential between  baryons a  la Born-Oppenheimer
%% \cite{suganuma}. This is, however, not applicable to the nucleons with
%% light quarks.
There are several studies of the $NN$ interactions on the lattice, which
do not rely on L\"uscher's method.  One uses the Born-Oppenheimer
picture,  i.e.,  if one  of  the three  quarks  inside  the baryon  is
infinitely heavy,  one may define  the potential between baryons  a la
Born-Oppenheimer \cite{suganuma}.  This is, however, not applicable to
the nucleons with light quarks.
The other employs  the strong coupling limit, which  has been proposed quite
recently \cite{forcrand}.
}
   Furthermore, it utilizes the
   finiteness of the lattice box effectively to extract the
  information of the on-shell scattering matrix and the 
  phase shift.  This approach has been applied to
   extract the $NN$ scattering lengths in the quenched
   QCD simulations \cite{Fukugita:1994ve} and in the 
  (2+1)-flavor QCD simulations with the mixed action \cite{NPLQCD}.
  
  Recently, the present authors proposed a closely related but an
   alternative
  approach to the $NN$ interactions from lattice QCD \cite{Ishii:2006ec,Aoki:2008hh}.  
  The starting point is the
  same BS wave function  $\psi (\br)$ as discussed in Ref.~\citen{luescher}.  
  Instead of looking at the wave function 
  outside the range of the interaction,
  we consider the internal region $ |\br | < R$ and
  define the energy-independent  non-local potential $U(\br, \br')$
  from $\psi (\br)$ so that  it   
  obeys the 
  Schr\"{o}dinger type equation in a finite box.
  Since $U(\br, \br')$ for strong interaction
  is localized in its spatial coordinates due to confinement
   of quarks and gluons, the potential receives
   finite volume effect only weakly in a large box. Therefore, 
   once $U$ is determined and is appropriately extrapolated to 
   $L \rightarrow \infty$, one may simply use the Schr\"{o}dinger
   equation in the infinite space to calculate the scattering phase shifts
    and bound state spectra to compare  with  experimental data.    
  Further advantage of utilizing the potential is that it would be a smooth
   function of the quark masses so that it is relatively easy to handle
  on the lattice. This is in sharp contrast to the 
   scattering length which shows a singular 
  behavior around the quark mass corresponding to the 
  formation of the $NN$ bound state.\footnote{
  Similar situation is well-studied in connection with the 
   BEC-BCS crossover in cold  fermionic atoms \cite{BEC-BCS}, where
   the external magnetic field plays a role of the quark mass in QCD. 
  For seminal suggestion on the rapid quark-mass dependence of the 
   $NN$ scattering length, see Ref.~\citen{Kuramashi:1995sc}.}

  Since we consider the non-asymptotic region ($ |\br | < R$)  
  of the wave function, the resultant potential $U$ and the $T$-matrix
  are  off-shell.  Therefore, they
   depend on the nucleon interpolating operator adopted to define
   the BS wave function.  This is in a sense  an advantage, since 
    one can establish a one-to-one correspondence between the nucleon
   interpolating operator and the $NN$ potential in QCD, which is not attainable
   in phenomenological $NN$ potentials.  It also implies that 
  the $NN$ potential on the lattice and the phenomenological $NN$ potentials
  are equivalent only in the sense that they give the same observables,
  so that the  
  comparison of their spatial structures should be made only qualitatively.
 
  The purpose of this paper is twofold:  First, we will present
  a theoretical foundation of our method to extract the $NN$ potentials from
  lattice QCD. Then, we will give a full account of the application of the 
  method to the quenched lattice QCD simulations.
     Once our method in lattice QCD is proved to work in the $NN$ system,
  it will   have various applications not only to nuclear many-body problems
  but also  to hyperon-nucleon, hyperon-hyperon and three-nucleon interactions
   which have much less experimental information than the $NN$ systems. 
  A first attempt to the hyperon-nucleon potential has been already
  reported in Ref.~\citen{Nemura:2008sp}, and more on hyperons will appear in the future publications. 
 
 This paper is organized as follows.
 In \S\ref{sec:strategy}, we illustrate the derivation of the two-body and many-body potentials from the 
 wave function in quantum mechanics.  
 In \S\ref{sec:BS-wave-spin_1/2}, the idea in the previous section is generalized to the 
  interaction of composite particles in field theory.
 In \S\ref{sec:OM-potential},  we classify the general structure  of the $NN$ potential in the velocity expansion and show  the procedure to determine each term.
 In \S\ref{sec:central}, the method to determine the $NN$ potential from the lattice QCD data is 
 discussed in detail for the effective central potential at low energy.
 We also discuss the method to extract  the tensor potential in our approach. 
In \S \ref{sec:numerical},  $NN$ potentials obtained from  the quenched lattice QCD simulations 
  are presented.
  Section \ref{sec:summary} is devoted to summary and concluding remarks.
 In Appendix \ref{sec:BS-infinite},   
 a   field-theoretical derivation of
  the asymptotic  BS wave function at large distance is presented.
   In Appendix \ref {sec:OM}, the way to make general decomposition of the  $NN$ potential (the Okubo-Marshak decomposition\cite{okubo})
    is reviewed. In Appendix \ref{sec:matrix_element},  matrix elements of the general $NN$ potential
     are presented. 
In Appendix D, heat-kernel representation of the Green's function is presented.

%%%%%%%%%%%%%%%%%%%%%%%%%%%%%%%%%%%%%%%%%%%%%%%%%%%%%%%%%%%%%%%%%%%%%%%%
\section{Non-local potential in quantum mechanics}
\label{sec:strategy}
%%%%%%%%%%%%%%%%%%%%%%%%%%%%%%%%%%%%%%%%%%%%%%%%%%%%%%%%%%%%%%%%%%%%%%%% 

\subsection{Two-body force}
\label{sec:2-body}

To show the basic concept of  the non-local potential in a finite box
 with the size $L \times L \times L$,
 we start with a non-relativistic two-body problem described 
 by the stationary Schr\"{o}dinger equation:
\beq
(\nabla^2 + k_n^2) \psi_n(\br) = 2\mu \int U(\br,\br') \psi_n(\br') d^3r' ,
\label{eq:QM_Schroedinger}
\eeq
 where 
  $\br$ is the relative coordinate of the two spinless and  non-relativistic
  particles, and $k_n$ is related to 
  the discrete energy eigenvalues  $E_n = k_n^2/(2\mu)$ 
($n=0, 1,2, \cdots)$ with $\mu$ being the reduced mass.
  The wave function  obeys
   the periodic boundary condition. The non-local potential 
   $U(\br,\br')$\footnote{Here
    we use the standard  term ``non-local"  in the sense that 
    $U(\br,\br')$ cannot be written as  $V(\br) \delta (\br - \br')$.}
     is assumed to be energy-independent and Hermitian, 
 $U^*(\br',\br)=U(\br,\br')$, so that the discrete energy eigenvalues $E_n$ are 
 real and corresponding eigenfunctions can be made orthonormal.
 For the scattering states (bound states) in the infinite volume,
 we have  $E (L\rightarrow \infty)>0$ 
  ($E (L\rightarrow \infty)< 0$). 
 On the other hand, negative $E_n(L)$ in the finite volume does not necessarily imply
  the existence of the bound state at $L\rightarrow \infty$.

    We consider the potential whose
  spatial extension  is sufficiently
 small in the sense that  $U(\br,\br')$ is exponentially suppressed 
 for $ \{ |\br|, |\br'| \}  > R$ with $R$ being
 smaller than $L/2$.  We define the ``inner region" by
 ${\Omega}_{\rm in}=\{ \br \in L^3 | \ |\br| < R \}$. 
  Then, the wave function in the ``outer region"
 $\Omega_{\rm out}=L^3-{\Omega}_{\rm in}$
 satisfies the Helmholtz equation, $(\nabla^2 + k_n^2) \psi_n(\br) =0$,
  with the periodic boundary condition.

Let us consider the following inverse problem:
Suppose we have no information about  $U$ except that
 it is  smooth and short ranged,
  while we know linearly independent
  wave functions $\psi_n(\br)$ and associated 
  energy  $E_n = k_n^2/(2\mu)$
   in a finite box  for  $n \le n_c$.\footnote{This is more luxurious situation
  than the usual inverse scattering problem where only the scattering
  phase shifts in the outer region are available.}   
 Now, we introduce the following  function:
 \beq
 K_n(\br)= \frac{1}{2\mu} (\nabla^2 + k_n^2) \psi_n(\br)
 = \langle \br | (E_n -H_0) |n \rangle ,
 \label{eq:kernel}
\eeq
 where $H_0$ is the non-relativistic kinetic energy operator satisfying
 $ \langle \br | H_0 |n \rangle = \linebreak  \frac{-1}{2\mu} \nabla^2  \psi_n(\br)$.
 Since  $(\nabla^2 + k_n^2)$ 
 removes the non-interacting part of the wave function, 
  $K_n(\br)$  is non-vanishing only in the inner 
 region $\Omega_{\rm in}$ irrespective  of the sign of $k_n^2$.
   
  By taking into account the fact that $\psi_n(\br)=\langle \br | n \rangle$ may not be
 orthonormal, we introduce  the norm kernel ${\cal N}_{n n'} \equiv \langle n|n' \rangle
   = \int d^3r \psi^*_{n}(\br) \psi_{n'}(\br)$, so that the projection 
    operator to the space spanned by the wave functions with $n \le n_c$ 
    reads ${P}(n_c)= \sum_{n,n'}^{n_c} |n \rangle {\cal N}^{-1}_{n n'} \langle n'| 
  \equiv \sum_{n}^{n_c} {P}_n  $.
 Then, an energy-independent and  non-local potential can be defined as 
 \beq
 U(\br,\br')  
= \langle \br | \left[ \sum_n^{n_c} (E_n-H_0 ) {P}_n \right] | \br' \rangle  
=  \sum_{n,n'}^{n_c} K_n(\br) {\cal N}^{-1}_{n n'}  {\psi}^*_{n'}(\br') ,
\label{eq:QM_potential}   
 \eeq
 which leads to the Schr\"{o}dinger equation Eq.~(\ref{eq:QM_Schroedinger})
  for $\psi_{n \le n_c} (\br)$. If we apply a unitary transformation $A$ to the 
   wave function, $\psi \rightarrow \psi' =A \psi$, 
   the non-local potential is modified as  $U \rightarrow U' = A U A^{\dagger}$.  
   Such unitary transformation does not affect the observables, while it
   changes the spatial structure of the wave function and the non-local potential.

   If $E_n$ are all real and  ${\cal N}_{n n'}=\delta_{nn'}$,
  the potential ${U} =  \sum_n^{n_c} (E_n-H_0) {P}_n  $ becomes a
  hermitian operator   $\langle n  | U | n'  \rangle^* = \langle n'  | U | n  \rangle $
  in the subspace $n \le n_c$.
    Otherwise, the hermiticity is not
  obvious and should be checked case by case.
  In field theory discussed later,  $\psi_{n}(\br)$ corresponds to the 
  equal-time Bethe-Salpeter amplitude in a finite box and $E_{n_c}$
  corresponds to the threshold energy $E_{\rm th}$ of inelastic
    channels. 
 
  In practice, the potential defined in Eq.~(\ref{eq:QM_potential}) 
  has  limited use, because  the 
   number of states satisfying the condition  $E \le E_{\rm th}$ is not 
   generally large for lattice QCD in a finite box.
     This problem can be evaded when we focus on the 
  low-energy scattering with $E$ sufficiently smaller than
   the intrinsic scale of the system or the 
    scale of the non-locality of the potential.   
   In such a case,  the velocity expansion
   of $U(\br,\br')$ in terms of its  non-locality is useful \cite{TW67}:
  For example, a  spin-independent potential with hermiticity, 
   rotational invariance, parity symmetry,
   and time-reversal invariance can be expanded as 
 \beq
\label{eq:U-del}
 \! \! \!   U(\br,\br')    &=& V(\br, \bv) \delta(\br-\br'), \\
 \! \! \!   V(\br, \bv) &=& V_0(r) + \frac{1}{2} \{ V_{v^2}(r), \bv^2 \}   
                       + V_{\ell^2}(r) \bL^2 + \cdots ,
 \label{eq:V-pot}
 \eeq  
   where $\bv = \bp/\mu $ and $\bL = \br \times \bp $ with 
   $\bp = -i \nabla$. 
 Each coefficient of the expansion is  
 the local potential and  can be determined  successively 
  by the wave functions  at low energies:  
  For example, if we have five wave functions corresponding to
  $E_{n=0,1,2,3,4}$, we obtain 
 \beq
 (E_n-H_0) \psi_{n}(\br)  =
  \left[  V_0(r) + \frac{1}{2} \{ V_{v^2}(r), \bv^2 \}  
         + V_{\ell^2}(r) \bL^2 \right] \psi_{n}(\br) .
 \label{eq:V-algebra}   
 \eeq
 Pretending that $V_{v^2}(r)$ and $(\frac{\partial}{\partial r})^n V_{v^2}(r)$
 are independent of each other,
  Eq.~(\ref{eq:V-algebra}) for $n=0, \cdots, 4$ can be solved algebraically to obtain
 $V_0(r), V_{v^2}(r), \frac{\partial}{\partial r} V_{v^2}(r),
 (\frac{\partial}{\partial r})^2 V_{v^2}(r)$ and $V_{\ell^2}(r)$.
 Hermiticity of the potential can be checked  by
 the consistency among the local potentials thus determined.
  Stability of the potentials against the number of 
  wave functions introduced  can be also checked. 

 An advantage of defining the potential 
 from the wave functions in the ``inner region" is that
 the effect of the periodic boundary condition is 
 exponentially suppressed for finite range interactions:
  Then one can first make 
  appropriate extrapolation of
     $U(\br,\br')$ or $V(\br, \bv)$  to $L\rightarrow \infty$, and then
 solve the Schr\"{o}dinger equation  using the extrapolated potential
   to calculate the observables such as 
    the phase shifts and binding energies in the 
    infinite volume.\footnote{Strictly speaking, the local potentials
    with higher derivatives must be treated as  perturbation
    to keep the Schr\"{o}dinger equation as a second order differential equation.}
   This is in contrast to the approach
   by  L\"{u}scher \cite{luescher} in
   which  the wave functions in the ``outer region" suffering from 
   the boundary conditions is ingeniously utilized to probe the 
  scattering observables.  Apparently, 
  the two approaches are  the opposite
  sides of a same coin.
 
\subsection{Many-body forces}
\label{sec:many-body}

 For the interactions among composite 
 particles, there are in principle many-body forces which take place
 in the system composed of more than two particles.
 The well-known example in nuclear physics is the Fujita-Miyazawa type
  three-body force acting among three nucleons 
  \cite{Fujita:1957zz,Weinberg:1992yk}. It is phenomenologically important
   for the extra binding of light nuclei \cite{Pieper:2001ap} and for the 
   extra repulsion in high density matter \cite{Akmal:1998cf} and in elastic
    nucleus-nucleus
    scatterings \cite{Furumoto:2009zz}.
    
The method to define the two-body potential from the relative wave function
discussed above can be generalized to the many-body forces.
 Let us illustrate the procedure by considering
  the three-body system of spinless and
 distinguishable particles with equal mass $m$.
 We consider the local potentials for both
  two-body and three-body forces just for simplicity. 
 In the rest frame of the three-body system, 
 we have
\beq
(E_n - H_{0r} - H_{0 \rho} ) \psi_n(\br, \brho) = 
 \left[ \sum_{i>j}V_2(\bx_i, \bx_j) + V_{3} (\bx_1, \bx_2, \bx_3) \right]
 \psi_n(\br, \brho),
\label{eq:QM_Schroedinger-3}
\eeq
 where $\br (= \bx_1 - \bx_2)$ and $\rho (=\bx_3 -(\bx_1+\bx_2)/2)$ 
 are the Jocobi coordinates.  
 $H_{0r} =- \nabla_r^2/(2\mu_r)$ and
 $H_{0\rho} = - \nabla_{\rho}^2/(2\mu_{\rho})$
 are the kinetic energy operator  with $\mu_r (=m/2)$ and 
 $\mu_{\rho} (=2m/3)$ being the reduced masses. $E_n$ is the
  total energy of the three-body system at rest.
  Because of the translational invariance, the two-body 
   potential $V_2$ and the three-body potential
    $V_3$ are the functions of $\br$ and $\brho$. 

 If we know the wave function and the total energy on the 
  left-hand side of Eq.~(\ref{eq:QM_Schroedinger-3}), the three-body 
  potential can be determined by the following procedure.
 We first consider the situation, $|\brho| \gg  R \gg |\br|$, where
 $V_2(\bx_2,\bx_3)$, $V_2(\bx_1,\bx_3)$ and
 $V_3(\bx_1,\bx_2,\bx_3)$ are vanishingly small 
  because of the assumed short-range nature of the 
  potentials.  Then,
  $V_2(\bx_1,\bx_2)$ can be determined by changing $\br$ within
   the range $R > |\br|$. One can carry out similar
    procedure to determine $V_2(\bx_2,\bx_3)$ and $V_2(\bx_1,\bx_3)$.
 Alternatively, one may determine $V_2$ from the 
   genuine two-body system.
   
  Once all the two-body potentials are determined, $V_3$ can be
   extracted from the wave function in the range,  $R > |\br|$ and 
   $R > |\brho|$,  through  the three-body equation
   Eq.~(\ref{eq:QM_Schroedinger-3}).
    It is important to note that
   the three-body potential is always obtained together
    with the two-body potential: they are closely tied 
   through the wave function.
    If one makes the unitary transformation of the wave function,
    both $V_2$ and $V_3$ are changed simultaneously.
    
  The above procedure can be formally generalized to 
   the non-local potentials and  to the $N (> 3)$-particle systems with 
    different masses and internal degrees of freedom.

%%%%%%%%%%%%%%%%%%%%%%%%%%%%%%%%%%%%%%%%%%%%%%%%%%%%%%%%%%%%%%%%%%%%%%%%%%%%%%%
\section{Non-local potential in field theory for spin 1/2 particles}
\label{sec:BS-wave-spin_1/2}
%%%%%%%%%%%%%%%%%%%%%%%%%%%%%%%%%%%%%%%%%%%%%%%%%%%%%%%%%%%%%%%%%%%%%%%% 

\subsection{Bethe-Salpeter wave function}

 In field theory, the best analogue of the two-particle wave function 
 is the equal-time Bethe-Salpeter (BS) amplitude, so that 
 we use the term ``BS wave function" throughout this paper. 
 Let us consider the following BS wave function
 for the 6-quark state with total energy $W$ and the total three-momentum
  $\bP=0$ in a finite box  $L^3$,
\begin{eqnarray}
\Psi_{\alpha \beta}(\br,t)  = 
\langle 0 \vert  {n}_{\beta}(\by,t) {p}_{\alpha}(\bx,t) \vert B=2;W,\bP=0 \rangle \equiv \psi_{\alpha \beta}(\br) e^{-iWt},
\label{eq:BS-def-def}
\end{eqnarray}   
where the relative coordinate is denoted as $\br = \bx - \by$. 
The   local composite operators for the proton and the neutron
 are denoted by   $p_{\alpha}(\bx,t)$ and  $n_{\beta}(\by,t)$
 with spinor indices $\alpha$ and $\beta$.
  The QCD vacuum is denoted by $| 0 \rangle $, while the
 state $|{B}=2 ; W,\bP=0  \rangle $ is a
 QCD eigenstate with baryon number 2 and with the same quantum numbers as the pn system.
  One should keep in mind  that $|{B}=2 ; W,\bP=0  \rangle $ is {\it not} 
 a simple superposition of a product state $| {\rm p} \rangle \otimes
|{\rm n} \rangle$,
 since there are complicated exchanges of quarks and gluons between the two composite
 particles.
  The stationary BS wave function $\psi(\br)$ 
 may be regarded as a probability amplitude in 
  $|{B}=2 ; W,\bP=0  \rangle $  to have 
 ``neutron-like'' three-quarks located at point $\by$ and
 ``proton-like'' three-quarks located at point $\bx$.

 The spatial extent of the $NN$ interaction in QCD is short ranged and is 
 exponentially suppressed beyond the distance $R > 2 $ fm.
 Therefore, the spatial part of the BS wave function in the 
  ``outer region" ($r > R$)
  satisfies the Helmholtz equation, 
  $((W/2)^2 -\nabla^2 + m_N^2 )\psi_{\alpha \beta}(\br) 
  =-( \nabla^2 + k^2 )\psi_{\alpha \beta}(\br) =0$,
 up to an exponentially small correction.
 Here the ``asymptotic momentum" $k$ is 
related to the  total energy  $W$ through the relation,  $W=2 \sqrt{k^2+m_N^2}$.
     To make a formal resemblance with the 
 non-relativistic case, we introduce  the ``effective center of mass  energy",
 $E=k^2/(2\mu)=k^2/m_N$ \cite{luescher}. 
  As shown in Appendix A, using the unitarity of the $S$-matrix, we can show that  the asymptotic behaviour of the BS wave function at large $r$ is identical to that of the scattering wave in the quantum mechanics, with the identification that the phase of the $S$-matrix is the scattering phase shift of the BS wave function.

  Now, we apply the same logic as the quantum mechanical case 
  in \S\ref{sec:2-body}.  The threshold of the pion production
  $E_{\rm th} \simeq m_{\pi} $  is chosen to be $E_{n_c}$. Namely,
  $(\nabla^2 + k^2) \psi_{\alpha \beta,E}(\br)$ is a function
   which has a support only in the inner region as long as 
    $E$ stays below the threshold.
   Thus we can define the short-ranged non-local potential as
 \beq
 \label{eq:QCD_Schroedinger}
 \! \! \! \! \! \! (E- H_0) \psi_{\alpha \beta,E}(\br) 
 &=&  \int U_{\alpha \beta; \gamma \delta}(\br,\br') \psi_{\gamma \delta,E}(\br') d^3r' , \\
 \label{eq:QCD_nonlocal-potential}
\! \! \! \! \! \!  U_{\alpha \beta; \gamma \delta}(\br,\br') 
 &=& \sum_{E,E'}^{E_{\rm th}} K_{\alpha \beta,E}(\br){\cal N}^{-1}_{EE'}
  {\psi}_{\gamma \delta,E'}^*(\br') \\
 \label{eq:QCD_local-potential}
\! \! \! \! \! \!  &=& V_{\alpha \beta; \gamma \delta}(\br, \bv) \delta(\br-\br') , 
\eeq  
 where $E=k^2/m_N$ and $H_0= -\nabla^2/m_N$. 
  By construction, the solution of Eq.~(\ref{eq:QCD_Schroedinger})  with 
  $U_{\alpha \beta; \gamma \delta}(\br,\br')$  
   extrapolated to $L \rightarrow \infty$
  reproduces the correct BS wave function in the asymptotic region, and hence the
   phase shifts and binding energies of the two-nucleon system.
    
 The Schr\"{o}dinger type equation with the non-local potential similar to
 Eq.~(\ref{eq:QCD_Schroedinger}) has been derived for bosons 
 on the basis of a diagrammatic method in Refs.~\citen{luescher} and
 \citen{luescher-CMP}.
 A slight difference is that our non-local potential
 has no explicit $E$-dependence by construction as seen 
  in Eq.~(\ref{eq:QCD_nonlocal-potential}).

\subsection{Interpolating operators}

In Eq.~(\ref{eq:BS-def-def}),
 simplest interpolating operators  for the neutron and 
the proton written in terms of the up-quark field
 $u(x)$ and the down-quark fields $d(x)$ would be 
\begin{eqnarray} 
   n_\beta(y) &=&
  \varepsilon_{abc} \left(
		     u_a(y) C \gamma_5 d_b(y)
		    \right)
  d_{c\beta}(y),
  \label{eq:neutron_op} \\
    p_\alpha(x) &=&
  \varepsilon_{abc} \left(
		     u_a(x) C \gamma_5 d_b(x)
		    \right)
  u_{c\alpha}(x),
  \label{eq:proton_op}
\label{eq:BS-def}
\end{eqnarray}   
where $x=(\bx, t)$, $y=(\by, t)$ and  
 the color indices are denoted by $a$, $b$ and $c$. The
 charge conjugation matrix in the spinor space is denoted
 by $C$.

  As shown in Appendix A, local operators such as given
   in Eqs.~(\ref{eq:neutron_op}) and (\ref{eq:proton_op}) are
   most convenient for relating  the BS wave function to the 
   four-point Green's function and the scattering 
   observables at $L \rightarrow \infty$. 
  Closely related observation was obtained 
  long time ago by Nishijima, Zimmermann and Hagg
  who derived the generalized  reduction formula   for 
  local composite fields \cite{NZ}. 

  In principle,  one may choose any composite operators
  with the same quantum numbers as the nucleon to define the BS wave
  function.\footnote{In practice, however, we had better restrict 
  ourselves to
  consider only {\it local} composite operators for the nucleon, since it is very 
  difficult, although not entirely impossible,
   to derive the reduction formula for non-local 
  composite operators without violating
the causality of relativistic theories.}
  Different  operators
  give different BS wave functions and different $NN$ potentials, although
  they lead to the same observables such as the phase shifts and binding energies.
  This is quite analogous to the  situation in quantum mechanics
  that the unitary transformation 
  of the wave function changes the structure of the 
   potential while  the observables are not modified. 
   A theoretical advantage of our approach based on lattice QCD is that
  we can unambiguously trace the one-to-one correspondence between
  the $NN$ potential and the interpolating operator in QCD.  
  This is in contrast to the phenomenological $NN$ potentials where
  connections to QCD operators are not attainable.

%%%%%%%%%%%%%%%%%%%%%%%%%%%%%%%%%%%%%%%%%%%%%%%%%%%%%%%%%%%%%%%%%%%%%%%%%%%%%%
\section{General form of the $NN$ potential}
\label{sec:OM-potential}
%%%%%%%%%%%%%%%%%%%%%%%%%%%%%%%%%%%%%%%%%%%%%%%%%%%%%%%%%%%%%%%%%%%%%%%%%%%%%%%

 In the previous section, we illustrated the 
 procedure to define the potential between the neutron and the proton, which
  has spinor indices $\alpha,\beta,\gamma,\delta$ running from 1 to 4.
In order to derive the general structure of the $NN$ potential
 at low energies, 
we restrict ourselves to consider only the upper components of these spinor indices 
  in the following sections.

\subsection{Symmetry of the two nucleon system}
 
\begin{table}[t]
\label{tab:states}
 \caption{Two-nucleon asymptotic states classified by the 
  total isospin $I$,  the total spin ($s$),  the orbital  angular momentum ($\ell$), 
   and  the total angular momentum ($J$) together with some examples in
    low partial waves. }
 \begin{center}
  \begin{tabular}{|c|c|c|c|c|c|c|}
    \hline
 $I$    & \multicolumn{3}{|c|}{0} &  \multicolumn{3}{|c|}{1} \\ 
    \hline \hline
 $s$    & 0 & \multicolumn{2}{|c|}{1}   & 0 & \multicolumn{2}{|c|}{1}   \\
    \hline
 $\ell$    & odd  & \multicolumn{2}{|c|}{even} & even & \multicolumn{2}{|c|}{odd}  \\
    \hline
 $J$    & $\ell$ &  $\ell$ & $\ell \pm 1$  & $\ell$  & $\ell$  & $\ell \pm 1$    \\ 
  \hline \hline
\rule[-2mm]{0pt}{6mm} $J=0$  & $-$ & $-$ & $-$ &   $^1{\rm S}_0$   & $-$ & $^3{\rm P}_0$    \\[1mm]
    \hline
\rule[-2mm]{0pt}{6mm} $J=1$  & $^1{\rm P}_1$  & $-$ & $^3{\rm S}_1$, $^3{\rm D}_1$   & $-$    &  $^3{\rm P}_1$ & $-$    \\
    \hline
\rule[-2mm]{0pt}{6mm} $J=2$  & $-$ & $^3{\rm D}_2$  & $-$ &   $^1{\rm D}_2$    & $-$ &  $^3{\rm P}_2$, $^3{\rm F}_2$   \\
    \hline
\rule[-2mm]{0pt}{6mm} $J=3$  & $^1{\rm F}_3$  & $-$ & $^3{\rm D}_3$, $^3{\rm G}_3$   & $-$    &  $^3{\rm F}_3$ & $-$    \\
   \hline
\rule[-2mm]{0pt}{6mm} $J=4$  & $-$  & $^3{\rm G}_4$ & $-$   & $^1{\rm G}_4$    &  $-$ & $^3{\rm F}_4$, $^3{\rm H}_4$    \\
    \hline
 \vdots & \vdots   & \vdots & \vdots &  \vdots & \vdots &  \vdots  \\[1mm]
    \hline
  \end{tabular}
 \end{center}
\end{table}

 It is useful to classify the asymptotic two-particle
 states by 
  the orbital  angular momentum ($\ell$), the total spin ($s$) and 
   the total angular momentum ($J$) together with the 
    total isospin $I$.  Using  
   the standard notation, $^{2s+1}\ell_J$, and taking into account 
   constraints due to Pauli principle, we have the 
    well-known relations given in Table I.

\subsection{Okubo-Marshak decomposition}

 The general form of the $NN$ potential in the two-component 
 spinor space has been classified by Okubo and Marshak\cite{okubo}. 
 We leave the derivation
  in Appendix B and recapitulate only the results here. 
 By using the helmiticity, translational invariance in space and time,
  Galilei invariance, rotational invariance, parity and time-reversal invariance,
  fermi statistics and isospin invariance, the potential has a general decomposition
\begin{eqnarray}
V &=& \sum_I V^{I}(\br, \bv, \bsigma_1, \bsigma_2) P^{\tau}_I ,\\
V^{I}&=& V^{I}_{0} + V^{I}_{\sigma}\, (\bsigma_1 \cdot \bsigma_2) 
   + \frac{1}{2} \{ V^{I}_{T},S_{12} \}  
    +   V^{I}_{LS}\,  {\bL} \cdot {\bS}
 + \frac{1}{2} \{ V^{I}_{P}, P_{12} \}
 + \frac{1}{2} \{ V^I_{W}, W_{12} \} ,
 \label{eq:general}\nonumber \\
\end{eqnarray}  
where $P_{\tau}^I$ is the projection operator to the iso-singlet ($I=0$) and 
iso-triplet $(I=1)$:
\beq
P^{\tau}_0=  \frac{1}{4}-\btau_1 \cdot \btau_2 ,
\ \ \
P^{\tau}_1 = \frac{3}{4}+ \btau_1 \cdot \btau_2 .
\eeq
Also, we define
\beq
S_{12} & = &  3(\bsigma_1 \cdot \hat{\br} )(\bsigma_2 \cdot \hat{\br} ) 
 - \bsigma_1 \cdot \bsigma_2 , \\
  {\bS}  &=&  \frac{1}{2} (\bsigma_1 + \bsigma_2), \ \ \ {\bL}  =  \br \times \bp , \\
P_{12}  & = & (\bsigma_1 \cdot \bv )(\bsigma_2 \cdot \bv ) , \\
W_{12}  & = & Q_{12} -\frac{1}{3} (\bsigma_1 \cdot \bsigma_2) {\bL}^2,\\
Q_{12}  & = & \frac{1}{2} \{ \bsigma_1 \cdot {\bL} , \bsigma_2 \cdot {\bL} \} ,
\eeq
with $\bv=\bp/\mu$. The anticommutators in Eq.~(\ref{eq:general})
  are necessary to make the potential hermitian, 
 since $S_{12},P_{12},W_{12}$
  do not commute with the scalar potentials 
    $V^I_A(\br^2, \bv^2, {\bL}^2)$ ($A=0, \sigma, T, LS, P, W$).

 If we keep the terms only up to the first order in $\bv$, 
  we obtain the conventional form of the potential at low energies
  commonly  used in nuclear physics:
\beq
V^{I}= V^{I}_{0}(r) + V^{I}_{\sigma}(r)\, (\bsigma_1 \cdot \bsigma_2) 
 + V^{I}_{T}(r)\, S_{12} +  V^{I}_{LS}(r)\, {\bL} \cdot {\bS}
 + O(\bv^2) ,
\label{eq:OM-pot}
\eeq
or in a more conventional notation,
\beq
\label{eq:OM-pot-2}
V & =&
   V_C(r) + V_T(r) S_{12} +  V_{LS}(r) {\bL} \cdot {\bS}  +{O}(\bv^2), \\
 & =&
   V_0(r)
  +V_\sigma(r)(\bsigma_1 \cdot \bsigma_2)
  +V_\tau(r)(\btau_1 \cdot \btau_2) 
   +V_{\sigma\tau}(r)
   (\bsigma_1 \cdot \bsigma_2)
   (\btau_1 \cdot \btau_2)
   \nonumber \\
 &&
  +\left[ V_{T0}(r)
  +V_{{T}\tau}(r)(\btau_1 \cdot \btau_2)\right] S_{12} \nonumber \\
 &&+\left[ V_{LS0}(r)
  +V_{{LS}\tau}(r)(\btau_1 \cdot \btau_2) \right] {\bL} \cdot {\bS}
  +{O}(\bv^2). 
 \label{eq:orderP}   
\eeq
The central and tensor potentials, $V_C$ and $V_T$,   in Eq.~(\ref{eq:OM-pot-2})
are the leading-order (LO) terms of $O(\bv^0)$ in the velocity expansion, while
 the spin-orbit potential, $V_{LS}$ is the next-to-leading-order (NLO) term
  of $O(\bv)$.

\subsection{Determination of the $NN$ potentials}

 For given $I$, $s$ and $J$, the matrix elements of 
 the LO and NLO potentials up to  $O(\bv)$
  in Eq.~(\ref{eq:orderP}) have the following structure (see Appendix C 
  and also see Ref.~\citen{TW67}):
\beq
&&\! \! \! \! \! \!  V^{I}(r;\ ^1\!J_J) = V_0^I(r) + V_{\sigma}^I(r) , \\
&& \! \! \! \! \! \!  V^{I}(r;\ ^3\!J_J) =  V_0^I(r) -3 V_{\sigma}^I(r)+ 2V_{T}^I(r)-V_{LS}^I(r) ,\\
&&\! \! \! \! \! \! V^{I}(r;\ ^3\!(J\mp 1)_J) = 
\left(
\begin{array}{cc}
V^{I}_{--}(r) & V_{-+}^I(r) \\
V_{+-}^I(r)   & V_{++}^I(r) \\
\end{array}
\right) ,
\eeq
with
\beq
V_{--}^I(r) &=& V_0^I(r)-3V_{\sigma}^I(r)
-\frac{2(J-1)}{2J+1}V_{T}^I(r)+(J-1)V_{LS}^I(r), \\
V_{++}^I(r) &=& V_0^I(r)-3V_{\sigma}^I(r) 
-\frac{2(J+2)}{2J+1}V_{T}^I(r)-(J+2)V_{LS}^I(r), \\
V_{-+}^I(r) &=& V_{+-}^I(r) = 6\frac{\sqrt{J(J+1)}}{2J+1} V_T^I(r) .
\eeq
 There are 8 unknown functions, $V^{I=0,1}_{0,\sigma,LS,T}$, while we have
 4 (2) diagonal and 1(0) off-diagonal matrix elements at each $J$ for  $J>0\ (J=0)$
  as seen from Table I.
 On the lattice, it is relatively  unambiguous
  to extract information for $\ell=0,1,2,3 = {\rm S, P, D, F}$
using  the irreducible representations of the 
cubic group \cite{luescher}. 
Then, at most 16 independent (14 diagonal and 2 off-diagonal)
 information as seen in Table I  are obtained for
 8 unknowns $V^I_A(r)$,
  so that each $V^I_A(r)$ can be determined in two different ways.
 
\subsection{Long range part of the potential}

 In QCD with dynamical quarks, the lightest hadron is the pion. Therefore,
 the longest range interaction between the nucleons is 
  dictated by the one-pion-exchange potential (OPEP).
  For later purpose, let us here summarize several features of OPEP with special 
  care about its chiral behavior.
  
First of all, the equivalence theorem implies that  the pseudo-scalar $\pi N$ coupling
 $g_{\pi N}(\simeq 14.0)$ and the pseudo-vector coupling $f_{\pi N}$ at low energy 
 are related through $f_{\pi N}= \frac{g_{\pi N}}{2M_N}$. This is simply obtained  by  
 kinematics.
 On the other hand, chiral symmetry leads to  the Goldberger-Treiman (GT) relation,
 $\frac{g_{\pi N}}{M_N} \simeq \frac{g_A}{F_{\pi}}$, where
  $g_A (\simeq  1.27)$ is the 
 nucleon axial-charge and $F_{\pi} (\simeq 93$ MeV) is the 
  pion decay constant. 

With these relations, the  OPEP reads
\beq
\label{eq:OPEP-1}
&&\hspace{-5mm}V_{\rm OPEP}(r)\nonumber\\
& = &  \frac{f_{\pi N}^2}{4\pi} ({\btau}_1 \cdot {\btau}_2)
({\bsigma}_1 \cdot {\nabla}_1)({\bsigma}_2 \cdot {\nabla}_2)
\frac{e^{-m_{\pi}r}}{r} \\ 
\label{eq:OPEP-2}
& = &  \frac{g_{\pi N}^2}{4\pi} \left( \frac{m_{\pi}}{2M_N} \right)^2
 \frac{({\btau}_1 \cdot {\btau}_2)}{3}
\left[ ({\bsigma}_1 \cdot {\bsigma}_2)
 + S_{12} \left( 1+ \frac{3}{m_{\pi}r} + \frac{3}{m_{\pi}^2 r^2} \right) \right]
\frac{e^{-m_{\pi}r}}{r}, \qquad\\
\label{eq:OPEP-3}
&=&   \frac{g_{A}^2}{4\pi} \left( \frac{m_{\pi}}{2F_{\pi}} \right)^2
 \frac{({\btau}_1 \cdot {\btau}_2)}{3}
\left[  ({\bsigma}_1 \cdot {\bsigma}_2)
 +  S_{12} \left( 1+ \frac{3}{m_{\pi}r} + \frac{3}{m_{\pi}^2 r^2} \right) \right]
\frac{e^{-m_{\pi}r}}{r} \\
\label{eq:OPEP-4}
& &  \xrightarrow[{\rm chiral\ limit}]{ }   \frac{g_A^2}{16\pi F_{\pi}^2}  
({\btau}_1 \cdot {\btau}_2)
\frac{S_{12}}{r^3} \ .
\eeq
Here we have used the equivalence theorem  to obtain Eq.~(\ref{eq:OPEP-2}) from 
Eq.~(\ref{eq:OPEP-1}) and use the GT relation to obtain 
Eq.~(\ref{eq:OPEP-3}) from  Eq.~(\ref{eq:OPEP-2}). $g_A$ and $F_{\pi}$ in 
Eq.~(\ref{eq:OPEP-4}) are  the values in the chiral limit.
 
 In quenched QCD without dynamical quarks,
  there arises a dipole ghost in the flavor-singlet channel (the $\eta$-channel in
   the case of two flavors) which couples to the nucleons \cite{Labrenz:1996jy,savage2}.
 The $\eta$-propagator in the quenched approximation
 is written as  
\beq
\label{eq:dipole-D}
D_{\eta}(q) 
= \frac{i}{q^2 - m_{\pi}^2 + i \epsilon } 
  +  \frac{iM_0^2(q)}{(q^2 - m_{\pi}^2 + i \epsilon )^2} ,
\eeq
where $M_0^2(q) \equiv m_0^2-\alpha_0 q^2$ with $m_0$ and $\alpha_0$ being
 ghost parameters.
The second term is the dipole ghost corresponding to the 
 hairpin diagram with quark-line disconnected.
Then the $NN$ potential from the $\eta$ exchange reads \cite{savage2}
%\newpage
\beq
&&\hspace{-5mm}V_{\eta}(r) \nonumber\\
& = &  \frac{f_{\eta N}^2}{4\pi} 
({\bsigma}_1 \cdot {\nabla})({\bsigma}_2 \cdot {\nabla})
\left[ (1-\alpha_0) + M_0^2(m_{\pi}) \frac{\partial}{\partial m_{\pi}^2} \right]
 \frac{e^{-m_{\pi}r}}{r} \\
&=&  \frac{g_{\eta N}^2}{4\pi} \left( \frac{m_{\pi}}{2M_N} \right)^2 
\frac{(1-\alpha_0)}{3}
  \left[({\bsigma}_1 \cdot {\bsigma}_2)
 + S_{12} \left( 1+ \frac{3}{m_{\pi}r} + \frac{3}{m_{\pi}^2 r^2} \right) \right]
  \frac{e^{-m_{\pi}r}}{r} \nonumber \\
& & - \frac{g_{\eta N}^2}{4\pi} \left( \frac{m_{\pi}}{2M_N} \right)^2 
\left( \frac{M_0^2(m_{\pi})}{2m_\pi} \right)  \frac{1}{3}
 \left[ 
 ({\bsigma}_1 \cdot {\bsigma}_2) \left( 1-\frac{2}{m_\pi r} \right)
 + S_{12} \left( 1+\frac{1}{m_\pi r}  \right)
  \right] e^{-m_{\pi}r} , \nonumber \\
\eeq
where $f_{\eta N}$ ($g_{\eta N}$) is
 the pseudo-vector (pseudo-scalar) coupling 
of the flavor-singlet $\eta$ to the nucleon. Its
 magnitude does not necessarily be as large as the $\pi N$ coupling
  \cite{Hatsuda:1989bi}.
 Note that
  the long range part of the potential has exponential fall-off instead
  of the Yukawa-type  because of the dipole-term in Eq.~(\ref{eq:dipole-D}).

Let us define  a ratio ${\cal R}_{\rm 13}$
between the central potential in the spin-singlet channel and that in the 
spin-triplet channel,
\beq
{\cal R}_{\rm 13} \equiv  \frac{V_{\rm C}(r;\ ^1S_0)}{V_{\rm C}(r;\ ^3S_1)} 
\xrightarrow[r \rightarrow \infty]{ }
    \left\{
  \begin{array}{c}
      +1  \ \ ({\rm one}{\mbox -}{\rm pion}{\mbox-}{\rm exchange}),\\
      -3  \ \ ({\rm one}{\mbox -}{\rm ghost}{\mbox -}{\rm exchange}). \\
  \end{array}
\right. 
\label{eq:ratio-F}
\eeq
Since we have $\langle {\bsigma}_1 \cdot {\bsigma}_2 \rangle_\textrm{spin-singlet} = -3$,
$\langle {\bsigma}_1 \cdot {\bsigma}_2 \rangle_\textrm{spin-triplet} = +1 $
 and the similar relations for the isospin, the large $r$ behavior of ${\cal R}_{\rm 13}$
  has different sign and magnitude between the one-ghost-exchange and one-pion-exchange.
  Therefore ${\cal R}_{\rm 13}$  can be  used  as a tool to identify the ghost contribution
   at large distance as will be discussed in \S \ref{sec:ghost}.

%%%%%%%%%%%%%%%%%%%%%%%%%%%%%%%%%%%%%%%%%%%%%%%%%%%%%%%%%%%%%%%%%%%%%%%%%%%%%%%
\section{Central and tensor forces in lattice QCD}
\label{sec:central}
%%%%%%%%%%%%%%%%%%%%%%%%%%%%%%%%%%%%%%%%%%%%%%%%%%%%%%%%%%%%%%%%%%%%%%%%%%%%%%%
 
\subsection{BS wave function on the lattice}

To define the BS wave function on the lattice  with the 
 lattice spacing $a$ and the spatial lattice volume $L^3$, 
 we start from the four-point correlator,
\begin{eqnarray}
\label{eq:4-point}
 {\cal G}_{\alpha \beta}(\bx, \by, t-t_0; J^P) 
&=& \left\langle 0
   \left|
    n_\beta(\by,t)
    p_\alpha(\bx,t)
    \overline{\cal J}_{pn}(t_0;J^P)
   \right| 0 
  \right\rangle  \\
&=& 
  \sum_{n=0}^{\infty} A_n 
  \left\langle 0
   \left|
    n_\beta(\by)
    p_\alpha(\bx)
    \right| E_n
  \right\rangle
  \ {e}^{-E_n(t-t_0)}, \\
&\xrightarrow[t\gg t_0] & 
\ \ A_0 \ \psi_{\alpha \beta}(\br;J^P)\  {e}^{-E_0(t-t_0)},
\label{eq:BSamp}
\end{eqnarray}
 with the matrix element $A_n=\langle E_n|\overline{\cal J}_{pn}(0)|0\rangle$. 
The states created by the source  $\overline{\cal J}_{pn}$
 have the conserved quantum numbers, 
   $(J,J_z)$ (total angular momentum and its z-component) and $P$ (parity).
 For studying the nuclear force in the  $J^P=0^+$ ($^1{\rm S}_0$) channel and
  the  $J^P=1^+$ ($^3{\rm S}_1$ and $^3{\rm D}_1$) channel, we adopt
   a wall source located at $t=t_0$ with 
    the Coulomb gauge fixing only at $t=t_0$:
\begin{eqnarray}
{\cal J}_{pn}(t_0;J^P)&=&
P_{\beta \alpha}^{(s)} \left[
p_{\alpha}^{\rm wall}(t_0)
n_{\beta}^{\rm wall}(t_0) \right] , 
\label{eq:wall-source}
\end{eqnarray}
where $p_{\alpha}^{\rm wall}(t_0)$ and
$n_{\beta}^{\rm wall}(t_0)$ are obtained by 
replacing the local quark fields $q(x)$ and $q(y)$ in
Eqs.~(\ref{eq:neutron_op}) and (\ref{eq:proton_op}) 
by the wall quark fields,
\begin{eqnarray}
 q^{\rm wall} (t_0) \equiv \sum_{\bxs} q(\bx,t_0).
\label{eq:wall-source2}
\end{eqnarray} 
By construction, the source operator Eq.~(\ref{eq:wall-source})
 has zero  orbital angular momentum at $t=t_0$, 
 so that states with fixed $(J,J_z)$ are obtained by
  the spin projection with $(s,s_z)=(J,J_z)$, e.g.
 $P^{(s=0)}_{\beta \alpha}=(\sigma_2)_{\beta \alpha}$
 and $P^{(s=1,s_z=0)}_{\beta \alpha}=(\sigma_1)_{\beta \alpha}$. 
Note that the $\ell$ and $s$ are not separately conserved: Therefore,
 the state created by the source ${\cal J}_{pn}(t_0;1^+)$ becomes a mixture
of the $\ell=0$ and $\ell =2$ at later time $t$.

 The BS wave function in the orbital S-state is then defined  with 
 the projection operator for the orbital angular momentum ($P^{(\ell)}$) 
 and that for the spin ($P^{(s)}$): 
 \begin{eqnarray}
\label{eq:1S0-wf}
\psi(r; ^1{\rm S}_0) &=& P^{(\ell =0)}  P^{(s=0)}   \psi(\br;0^+)
 \equiv {1\over 24} \sum_{g \in O}   P^{(s=0)}_{\beta \alpha}
   \psi_{\alpha \beta}(g^{-1}\br;0^+), \\
\label{eq:3S1-wf}
\psi(r; ^3{\rm S}_1) &=& P^{(\ell =0)}  P^{(s=1)}   \psi(\br;1^+)
 \equiv {1\over 24} \sum_{g \in O}   P^{(s=1)}_{\beta \alpha}
   \psi_{\alpha \beta}(g^{-1}\br;1^+).
\end{eqnarray}
Here the summation over $g \in O$ is taken for the cubic transformation 
group with 24 elements to project out the S-state.\footnote{
 More precisely, this projection picks up an $A_1^+$ state, which contains
not only an $\ell=0$ component but also the higher orbital waves with $\ell \ge 4$.
Latter contributions, however, are expected to be negligible at low energy. }$^,$\footnote 
{ Note that  $P^{(\ell =0)}  P^{(s=0)}$ in  Eq.~(\ref{eq:1S0-wf})
is a redundant operation, since we have already prepared $J^P=0^+$ state
 by the wall source ${\cal J}_{pn}(t_0;0^+)$ which allows only the $ ^1{\rm S}_0$ channel.
Also, $  P^{(s=1)}$ in Eq.~(\ref{eq:3S1-wf}) is  a redundant operation,
since  the  $J^P=1^+$  state   prepared  by  the  wall  source  ${\cal
J}_{pn}(t_0;1^+)$ allows only the spin-triplet state.
}

\subsection{Asymptotic momentum}
  
The asymptotic momentum $k$ for the S-states is  obtained 
 by fitting the BS wave function $\psi(\br)$ with the Green's function in a
finite and periodic box \cite{luescher}:
\begin{eqnarray}
\label{eq:GF}
 G(\br; k^2) = 
  \frac1{L^3}
  \sum_{\bns \in \boldsymbol{ Z}^3}
    \frac{e^{ i (2\pi/L) \bns \cdot  \brs }    }   { (2\pi/L)^2 {\bn}^2 - k^2},  
\end{eqnarray}
which  satisfies  $(\nabla^2  +  k^2)  G(\br;  k^2)  =  -  \delta_{\rm
lat}(\br)$   with   $\delta_{\rm   lat}(\br)$   being   the   periodic
delta-function.   In  the   actual  calculation,  Eq.~(\ref{eq:GF})  is
rewritten in terms  of the heat kernel ${\cal  K}$ satisfying the heat
equation, $\partial_t  {\cal K}(t, \br)  = \nabla^2 {\cal  K}(t, \br)$
with  the initial condition,  $ {\cal  K} (t  \rightarrow 0^+,  \br) =
\delta_{\rm  lat}(\br)$ (see  Appendix  \ref{sec:heat-kernel} for  the
detail).
  The fits are  performed outside the range  of the $NN$ interaction
   determined by  $\nabla^2\psi(\br)/\psi(\br)$ \cite{ishizuka}.

\subsection{Effective central potential at low energies}
 
 In the S-states at low energies, the effect of the 
 velocity dependent terms  in Eq.~(\ref{eq:orderP})
 is supposed to be small compared to the 
  velocity independent terms, so that  
   it is convenient to define  
 the  ``effective" central potential $V_{C}^{\rm eff}(r)$ \cite{Ishii:2006ec}: 
  \beq
  V_{C}^{\rm eff}(r) = 
  E + {1\over m_{\rm N}}{{\nabla}^2\psi(r)\over \psi(r)} .
\label{eq:naive_pot}
\eeq
 As long as we keep only the LO terms of the velocity expansion
  in Eq.~(\ref{eq:OM-pot-2}),
$V_{C}^{\rm eff}(r; ^1{\rm S}_0)$ is equivalent to $V_{C}(r; ^1{\rm S}_0)$,  while
$V_{C}^{\rm eff}(r; ^3{\rm S}_1)$ differs from 
$V_{C}(r; ^3{\rm S}_1)$ due to the higher order effects from the tensor potential.
 One can also study the validity of velocity expansion 
 in  Eq.~(\ref{eq:OM-pot-2}) by calculating  $V_{\rm C}^{\rm eff}(r; ^1{\rm S}_0)$
  for different energies $E$ (see \S \ref{sec:vel-dep}).

\subsection{Scattering lengths}

The $NN$ scattering lengths  for the S-states can be deduced from
L\"{u}scher's formula \cite{luescher,ishizuka},
\begin{equation}
 k \cot \delta_0(k) =
  \frac{2}{\sqrt{\pi} L} Z_{00}(1;q^2) =
  \frac{1}{a_0} + O(k^2), 
  \label{eq:SL}
\end{equation}
 where $Z_{00}(1;q^2)$ with $q=\frac{kL}{2\pi}$
 is obtained by the analytic continuation of the 
  generalized zeta-function $Z_{00}(s;q^2) =
   \frac{1}{ \sqrt{4\pi} } \sum_{{\bns} \in \boldsymbol{Z}^3}
    (\bn^2-q^2)^{-s}$ defined for ${\rm Re}\ s > 3/2$. 
 (See also Ref.~\citen{ishizuka2} for more general considerations.)
 In this formula, the sign of the S-wave scattering length 
 $a_0$ is defined to be positive for weak attraction.

\subsection{Decomposition into central and tensor potentials }
\label{sec:tensor}
 
 Although the tensor force at long distance is dominated by the 
 one-pion exchange, its spatial structure  at medium and short distances 
 is  not well understood  theoretically  nor well determined 
 phenomenologically. Therefore, it is quite important to extract it
  from lattice QCD.

 In the LO of the  velocity expansion in Eq.~(\ref{eq:OM-pot-2}),
only the central potential $V_C(r)$  and  the tensor potential  
$V_T(r)$ are relevant: The central potential  acts
separately on the S and D  components, while  the  tensor potential
 provides a coupling between these two. Therefore,
  we consider a coupled-channel
 Schr\"odinger equation in the  $J^P=1^+$ channel \cite{Ishii:2009zr}, in which 
 the BS wave  function  has  both S-wave
 and D-wave components:
\beq
  \bigl( H_0 + V_{C}(r) + V_{T}(r) S_{12} \bigr)
  \psi(\br;1^+)
  =
  E
  \psi(\br;1^+).
  \label{schrodinger.eq.one.plus}
\eeq 
The projections to the S-wave and D-wave components similar to 
  Eq.~(\ref{eq:3S1-wf}) read
\beq
{\cal P} \psi_{\alpha\beta} & \equiv &
  P^{(\ell =0)}  \psi_{\alpha \beta}(\br;1^+), \\
{\cal Q} \psi_{\alpha\beta} & \equiv &
 (1- P^{(\ell =0)} ) \psi_{\alpha \beta}(\br;1^+). 
\eeq  
 Note that both ${\cal P} \psi_{\alpha\beta}$ and ${\cal Q}\psi_{\alpha\beta}$ 
 contain additional components with $\ell\ge 4$ but they are expected to be small at low energies.

By multiplying ${\cal P}$  and ${\cal Q}$ to \Eq{schrodinger.eq.one.plus}
from the left and using the fact that  $H_0$,  $V_{C}(r)$ and $V_{T}(r)$ commute
with  ${\cal P}$ and  ${\cal Q}$,  \Eq{schrodinger.eq.one.plus} splits
into two equations,
\beq
  H_0 [{\cal P} \psi](\br)
  + V_C(r) [{\cal P}\psi](\br)
  + V_T(r) [{\cal P} S_{12} \psi](\br)
  &=&
  E
  [{\cal P} \psi](\br) ,
  \\
  H_0 [{\cal Q} \psi](\br)
  + V_C(r) [{\cal Q}\psi](\br)
  + V_T(r) [{\cal Q} S_{12} \psi](\br)
  &=&
  E
  [{\cal Q} \psi](\br),
\eeq
where we have suppressed the spin indices, $\alpha$ and $\beta$, for simplicity. 

 By picking  up  $(\alpha,\beta)=(2,1)$ component  of
these two  equations,  we arrive at 
\beq
  V_{C}(\br)
  &=&
  E
   - \frac1{\Delta(\br)}
  \big (
  [{\cal Q} S_{12}\psi]_{21}(\br) H_0 [{\cal P}\psi]_{21}(\br)
  -
  [{\cal P} S_{12}\psi]_{21}(\br) H_0 [{\cal Q}\psi]_{21}(\br)
  \big) ,
  \label{eq.vc}\qquad
  \\
  V_{T}(\br)
  &=&
  \frac1{\Delta(\br)}
  \big(
   [{\cal Q} \psi]_{21}(\br) H_0 [{\cal P} \psi]_{21}(\br)
  -
  [{\cal P} \psi]_{21}(\br) H_0 [{\cal Q} \psi]_{21}(\br)
  \big),\\
\Delta(\br) &\equiv & [{\cal  P}\psi]_{21}(\br) [{\cal  Q} S_{12}\psi]_{21}(\br) 
 - [{\cal Q}  \psi]_{21}(\br) [{\cal P}  S_{12} \psi]_{21}(\br).
\eeq

%%%%%%%%%%%%%%%%%%%%%%%%%%%%%%%%%%%%%%%%%%%
\section{Numerical results in quenched QCD}
\label{sec:numerical}

\subsection{Setup of the lattice simulations}
\label{sec:setup} 
  
   We  employ  the  standard  plaquette  gauge action on  a $32^4$  lattice  with
 the bare QCD coupling constant $\beta = 6/g^2 =  5.7$.
 The corresponding lattice spacing  is determined as  
 $1/a=1.44(2)$ GeV  ($a\simeq 0.137$ fm)  
  from the $\rho$ meson mass in the chiral limit \cite{kuramashi2}.
  The physical size of our lattice then reads $L\simeq 4.4$ fm.
   As for the fermion action, we adopt the standard Wilson quark action  with the 
  hopping parameter ($\kappa=0.1640, 0.1665$ and $0.1678$),
  which controls the quark masses.
 The  periodic  boundary  condition is  imposed on  the quark
  fields along the spatial direction, while
  the Dirichlet boundary  condition is imposed along the
temporal direction  on the time-slice  $t=0$.  The wall source is placed on
the time-slice at $t_0/a \equiv 5$ after the Coulomb gauge fixing at $t=t_0$.

  To generate the quenched gauge configurations,  
  we adopt the heatbath algorithm 
%with overrelaxation 
 and
 sample configurations are taken in every 200 sweeps after skipping
 3000 sweeps for thermalization. 
The  number of  sampled gauge configurations  $N_{\rm conf}$,  the  pion mass
$m_{\pi}$, the rho-meson mass $m_{\rho}$
   and the  nucleon   mass   $m_{ N}$   are  summarized   in
Table~\ref{tab:parameter}.  For $\kappa=0.1678$,  we have removed 
28 exceptional gauge configurations from the sample.

\begin{table}[t]
\caption{Summary of the hopping parameter $\kappa$,
 the pion mass $m_{\pi}$, the rho-meson mass $m_{\rho}$,
 the nucleon mass  $m_{\rm N}$, the time-slice  $(t-t_0)/a$ at
which  BS  wave  functions  are  extracted, the spatial-slice
$R/a$ above which the $NN$ potentials are inactive, and
the number  of gauge configurations $N_{\rm  conf}$ with exceptional configurations
being  removed.
 The lattice spacing is $a \simeq 0.137$ fm. Some numbers are updated 
 from Tables 1 and 2 of Ref.~\citen{Aoki:2008hh}. }
\begin{center}
\begin{tabular}{llllccc}
\hline\hline
$\kappa$ &  $m_{\pi}$ [MeV] & $m_{\rho}$ [MeV] 
& $m_{N}$ [MeV] & $(t-t_0)/a$ & $R/a$ & $N_{\rm conf}$ \\
\hline
0.1640 &  731.1(4) & 990.3(13) &  1558.4(63) & 7 & 11 & 1000 \\
0.1665 &  529.0(4) & 894.3(28) & 1333.8(82)  & 6 & 11 & 2000 \\
0.1678 &  379.7(9) & 837.9(21) & 1196.6(83)  & 5 & 12 & 2021 \\
\hline
\end{tabular}
\end{center}
\label{tab:parameter}
\vspace{0.5cm}
\end{table}

   The BS wave functions are measured
 at  $(t-t_0)/a  =  7, 6,  5$  
 for $\kappa=0.1640,  0.1665, \linebreak 0.1678$,  respectively. 
 These values of $t-t_0$ are determined by studying  the 
 ground state  saturation in the $NN$ potentials as discussed below. 
  We  employ the  nearest
 neighbor representation of  the discretized Laplacian as $\nabla^2
f(\bx) \equiv \sum_{i=1}^{3}\{ f(\bx + a\bn_i)\linebreak  + f(\bx -  a\bn_i)\} - 6  f(\bx)$, 
where $\bn_i$ denotes the
unit vector  along the $i$-th  coordinate axis. BS wave  functions are
fully measured for  $r <  0.7$ fm, where  rapid change of the
 $NN$ potential is expected. 
Since  the  change  is  rather  modest for  $r >  0.7$ fm,   the
measurement of BS wave functions has been restricted on the coordinate axes
and their  nearest neighbors to  reduce the %calculational  
 computational cost.

\subsection{BS wave functions in the S-state}

\begin{figure}[t]
\begin{center}
\includegraphics[width=7cm,angle=-90]{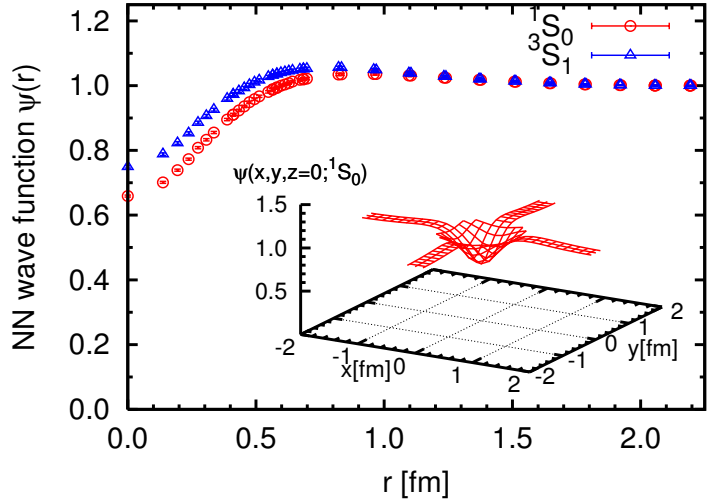}
\end{center}
\caption{The $NN$ wave functions in $^1{\rm S}_0$ and $^3{\rm S}_1$ channels
 for $m_{\pi}=529$ MeV ($\kappa=0.1665$).
The  inset  is a  three-dimensional
 plot of  the wave  function $\psi(x,y,z=0;\  ^1{\rm S}_0)$.
}
\label{fig1}
\end{figure}

Figure \ref{fig1} shows  the BS wave functions in  $^1{\rm S}_0$ and $^3{\rm S}_1$
channels for $\kappa=0.1665$.  The wave functions are normalized to be
 1 at the largest spatial point $r = 2.192$ fm. %for simplicity.

\begin{figure}[t]
\begin{center}
\includegraphics[width=5.5cm,angle=-90]{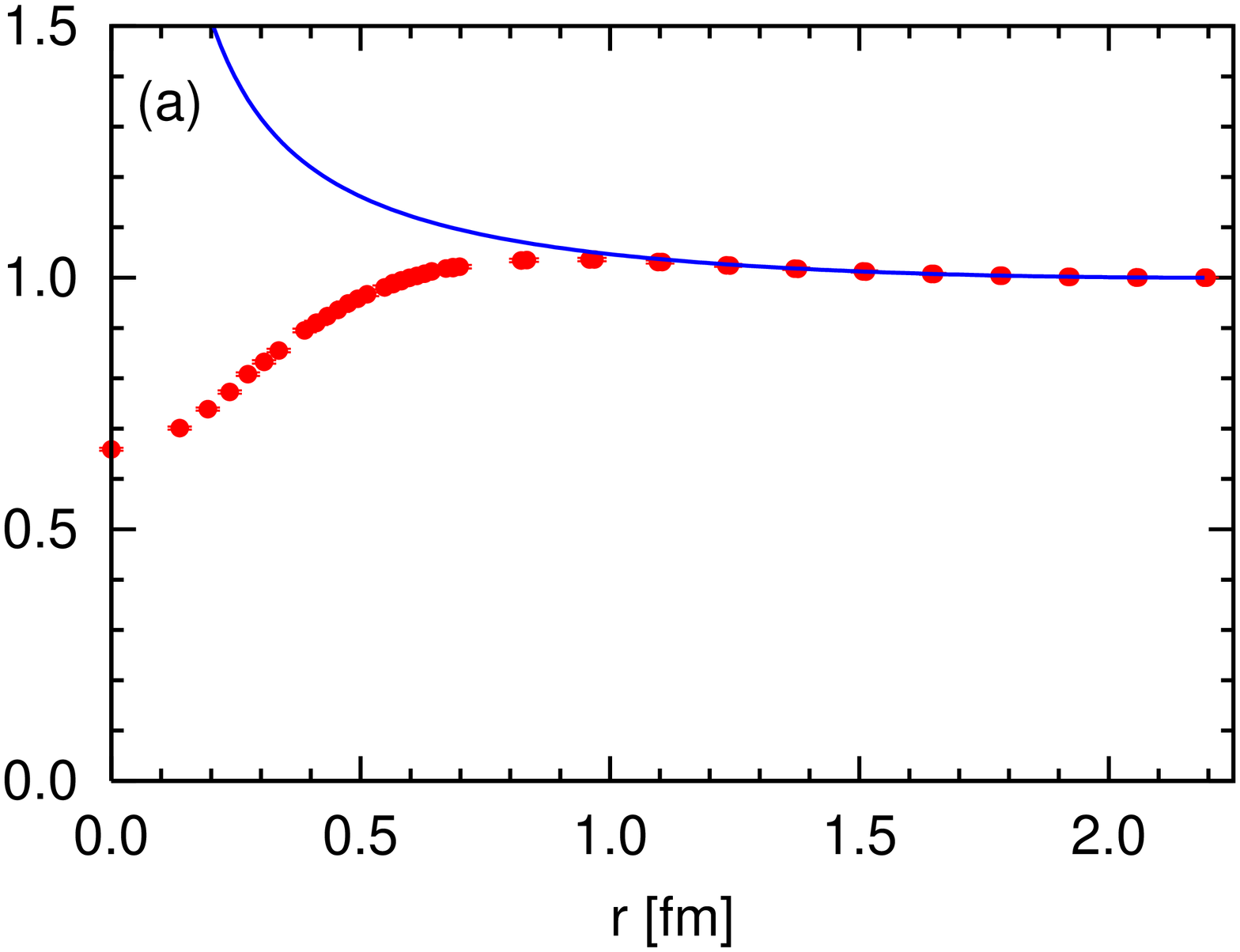}
\hspace{0.3cm}
\includegraphics[width=5.5cm,angle=-90]{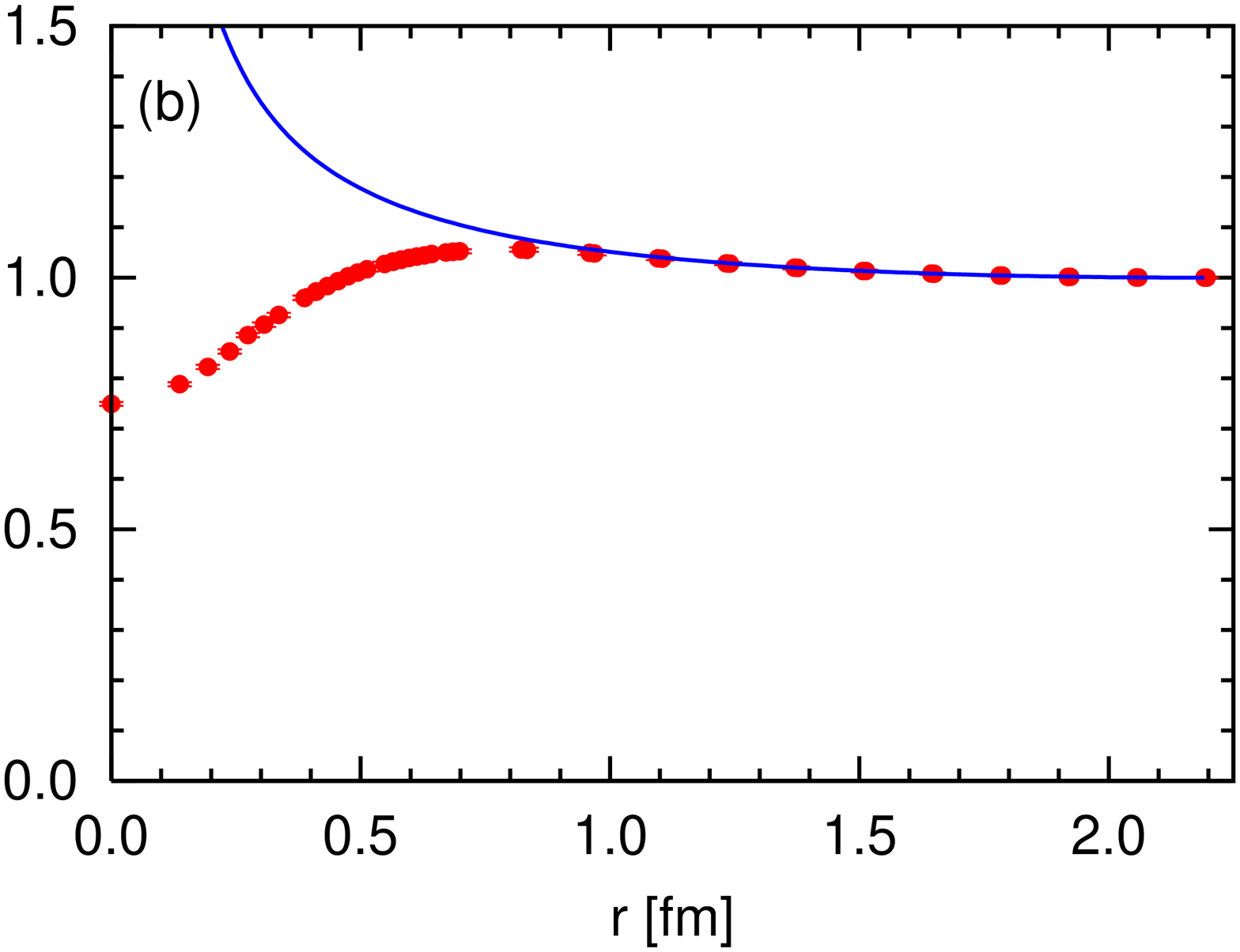}
\end{center}
\caption{(a) The fit of the  $NN$ wave functions 
 for $m_{\pi}=529$ MeV in the $^1{\rm S}_0$ channel using the
 Green's function in the  fit range $11 \le r/a \le 16$. 
(b) Similar fit for the $NN$ wave functions in the $^3{\rm S}_1$ channel.  
}
\label{fig2}
\end{figure}

Figure \ref{fig2}(a,b) show the fitting of the wave function
in the interval $R/a \le r/a \le 16$ using Eq.~(\ref{eq:GF}).
 This leads to the  values of  the  effective energy
$E \equiv k^2/m_{N}$ in Table \ref{tab:parameter}.
The value of $R$ is determined from the ground state saturation of the 
potential as discussed below.

\subsection{Effective central potential}
  
\begin{figure}[t]
\begin{center}
\includegraphics[width=8.5cm,angle=-90]{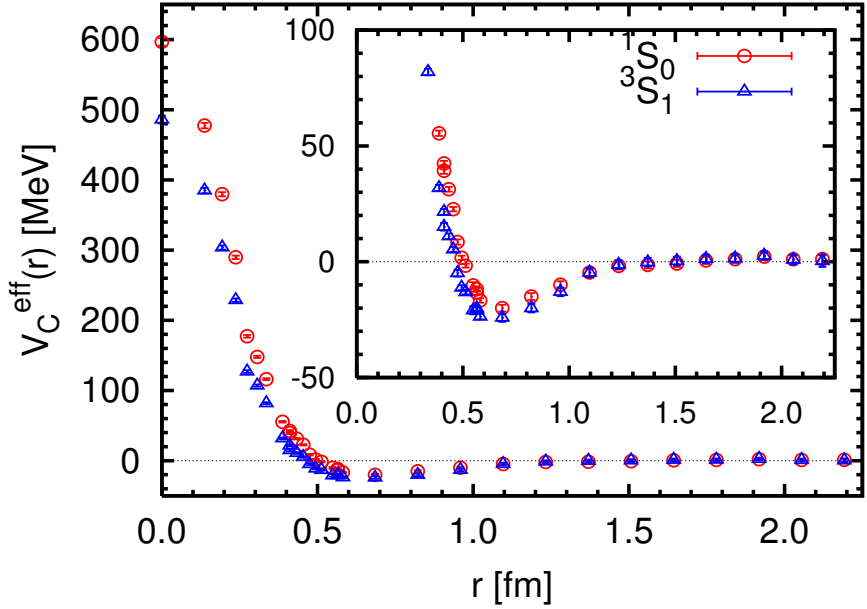}
\end{center}
\caption{
The effective central potentials in the $^1{\rm S}_0$ channel and 
 in the $^3{\rm S}_1$ channel for $m_{\pi}=529$ MeV. }
\label{fig3}
\end{figure}

\begin{figure}[t]
\begin{center}
\includegraphics[width=8.5cm,angle=-90]{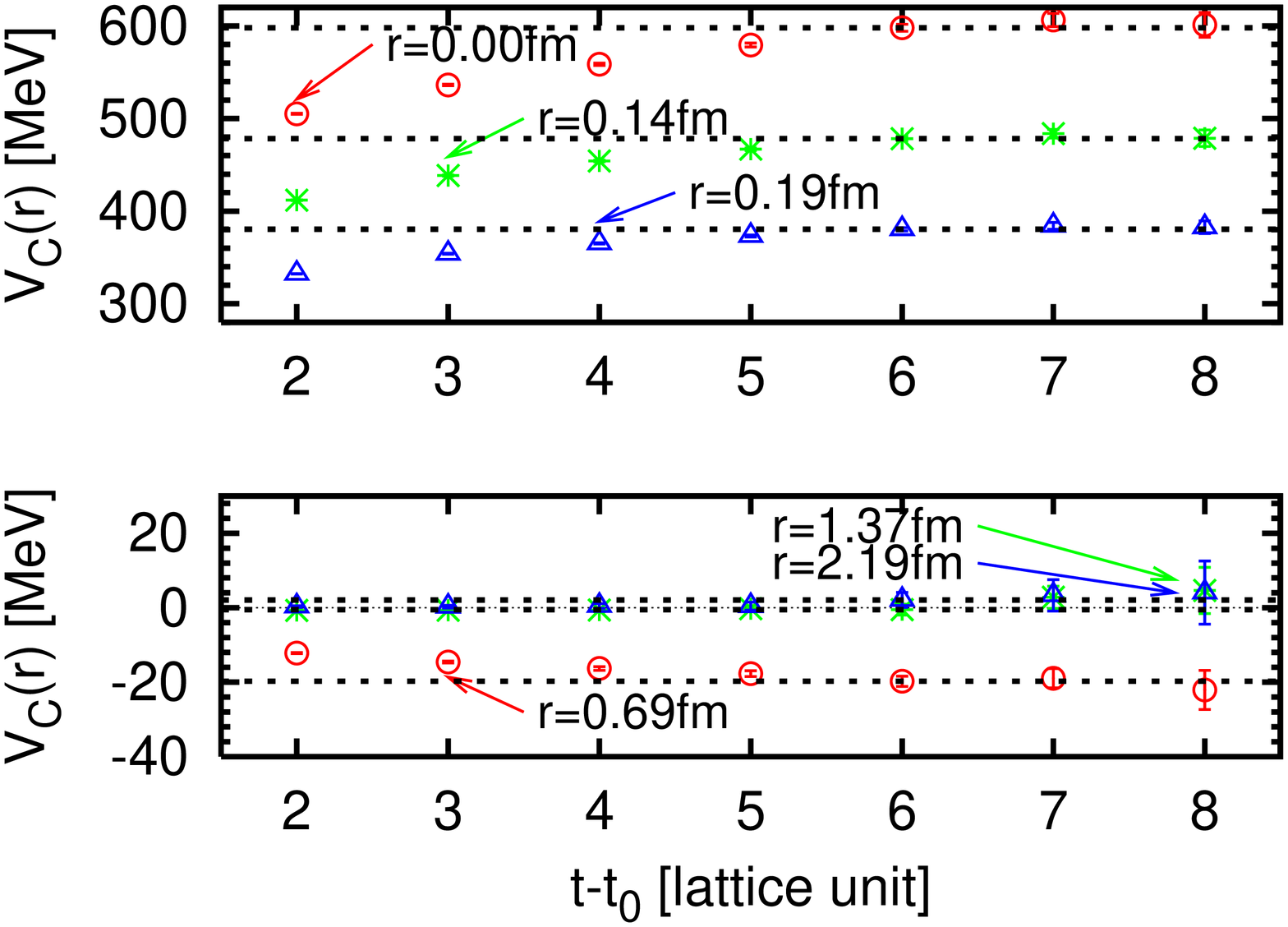}
\end{center}
\caption{
 The $t$-dependence of the potential at $r=0, 0.14, 0.19, 1.37, 2.19, 0.69$ fm from top to bottom
for the $^1{\rm S}_0$ channel at $m_\pi = 529$ MeV. 
}
\label{fig4}
\end{figure}
 
 Shown in Fig.~\ref{fig3} are the reconstructed  effective central 
 potentials in the $^1{\rm S}_0$ and $^3{\rm S}_1$ channels for $\kappa=0.1665$
  with the formula Eq.~(\ref{eq:naive_pot}).
 The overall structures of the potentials  are similar to the known phenomenological 
 $NN$  potentials discussed in 
  \S\ref{sec:intro}, namely
  the repulsive core at short distance
  surrounded by  the attractive well  at medium and long distances. 
  From this figure, we find that the interaction between the 
  nucleons is well switched
  off for $r > 1.5$ fm, so that we chose $R/a=11$ (for $m_{\pi}=731, 529$ MeV)
   and $R/a=12$ (for $m_{\pi}=380$ MeV) as given in Table \ref{tab:parameter}.
 { In both cases, the condition $R < L/2 = 2.2\ {\rm fm}$ is satisfied.}
      
  To check the stability of these potentials against the 
  time-slice adopted to define the BS wave functions, we 
   plot the $t$-dependence of the  $^1{\rm S}_0$ potential for several different values
 of $r$ as  shown in Fig.~\ref{fig4} for $m_{\pi}=529$ MeV:  In this
  case, choosing $(t-t_0)/a=6$ to extract $V_C(r)$ 
  would be good enough to assure the stability within the statistical errors.
  The time-slices chosen for other cases by the same procedure
  are given in Table \ref{tab:parameter}.

\subsection{Quark mass dependence of the central potential}

\begin{figure}[t]
\begin{center}
\includegraphics[width=8.5cm,angle=-90]{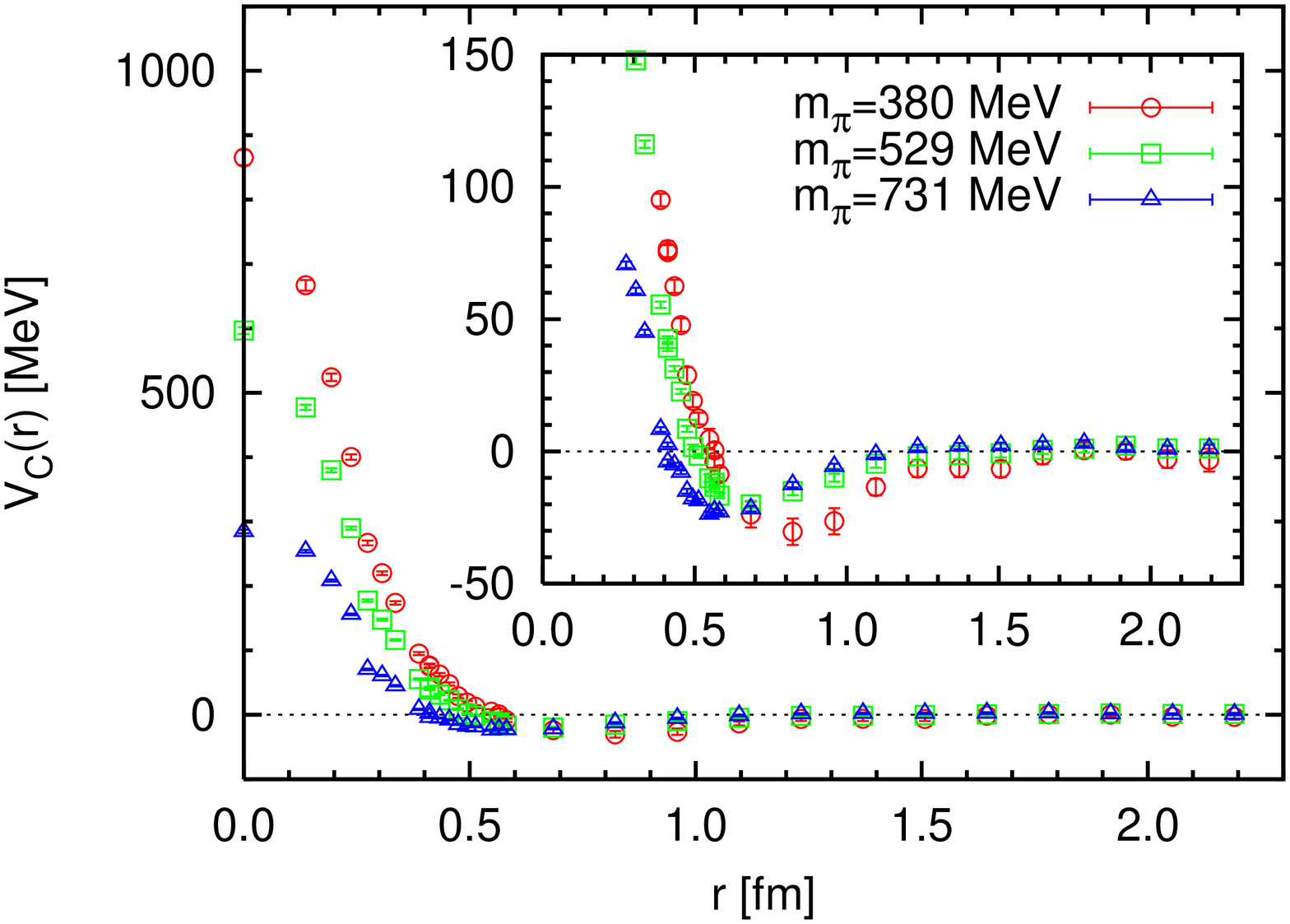}
\end{center}
\caption{
The central potentials in the $^1{\rm S}_0$ channel for three different quark masses. }
\label{fig5}
\end{figure}

\begin{figure}[t]
\begin{center}
\includegraphics[width=7cm,angle=-90]{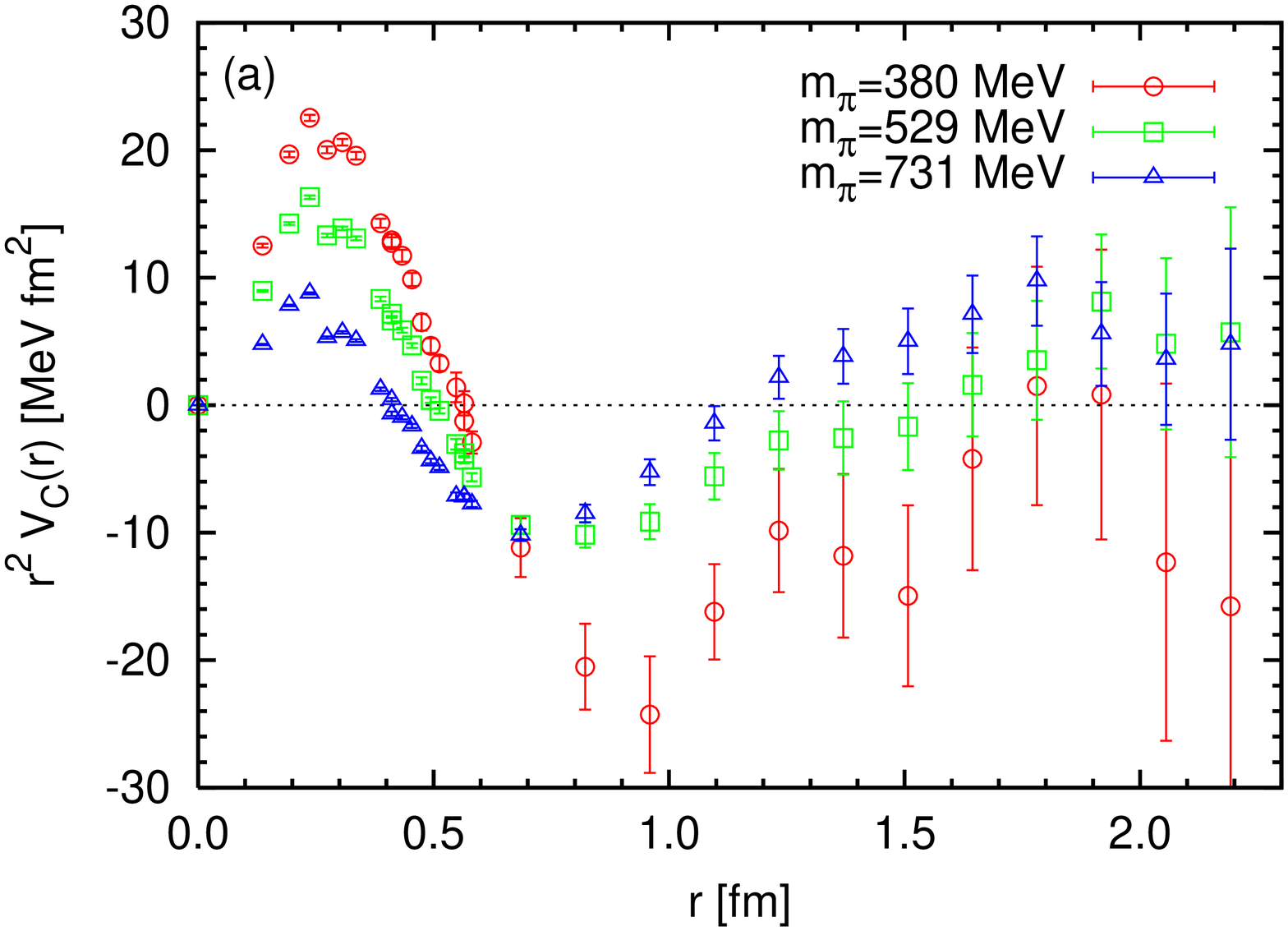}

\vspace{0.5cm}

\includegraphics[width=7cm,angle=-90]{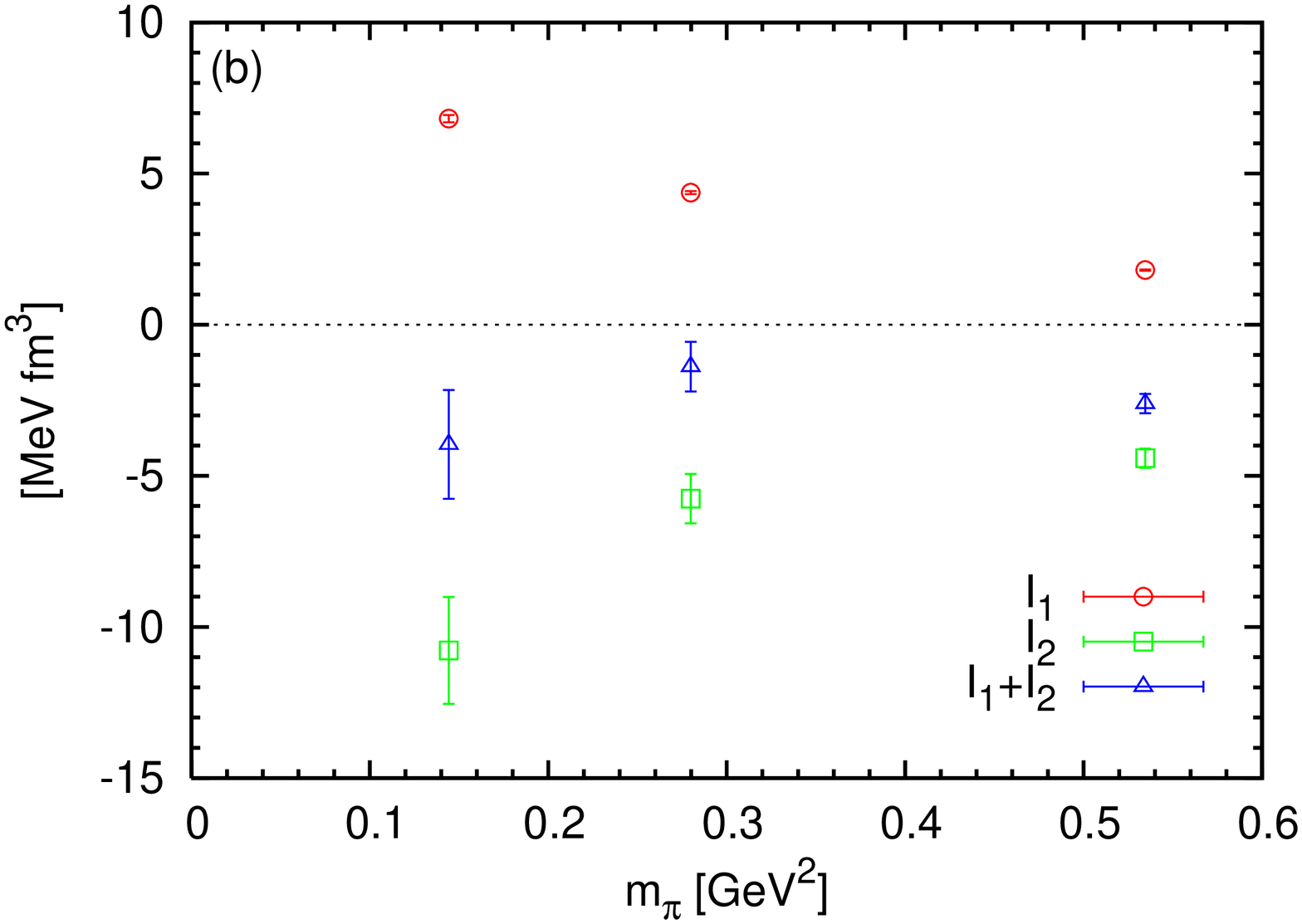}
\end{center}
\caption{(a)
The central potentials with $r^2$ multiplied 
in the $^1{\rm S}_0$ channel for three different quark masses. 
(b) Comparison of the attractive part and repulsive part of the 
 potential in terms of the volume integral in the $^1{\rm S}_0$ channel.}
\label{fig6}
\end{figure}
 
In Fig.~\ref{fig5}, we compare
 the  $NN$  central potentials  in  the $^1{\rm S}_0$ channel
  for  three  different quark  masses.
  As the quark  mass decreases, the repulsive core at short
 distance  and  the  attractive well  at  medium
distance are  enhanced simultaneously.  
 This feature  can be also seen in Fig.~\ref{fig6}(a)
 where  $r^2 V_C(r)$,   
 which appears in the quantum mechanical  matrix elements,
  is plotted.  To study the relative magnitude of the repulsion and the
   attraction, we define
   the following volume integrals of the potential and plot them
    in Fig.~\ref{fig6}(b):
\begin{eqnarray}
I_1= \int_0^{r_0} r^2 V_C(r) dr,  \ \ 
I_2= \int_{r_0}^{r_1} r^2 V_C(r) dr.
\label{eq:volume-int}
\end{eqnarray}
Here $r_0$ ($\sim$ 0.5 fm) is the first nodal point where 
 $r^2 V_C(r)$ changes sign from positive to negative,
  and  $r_1$ is the point at which $r^2 V_C(r)$   becomes essentially
  zero within the
  statistical errors.  The error bars in Fig.~\ref{fig6}(b) reflect
   the uncertainties of $r_{0,1}$ as well as those from the spline curve fit of the 
   data.   The comparison of $I_1$,
  $I_2$ and $I_1+I_2$ implies that (i) both repulsion and attraction
  increase in magnitude as quark mass decreases, and (ii) there is a large cancellation
   between the repulsion and attraction, and (iii) there is a net attraction increasing
    as the quark mass decreases.

\subsection{Dipole ghost in the central potential}
\label{sec:ghost}

\begin{figure}[t]
\begin{center}
\includegraphics[width=7cm,angle=-90]{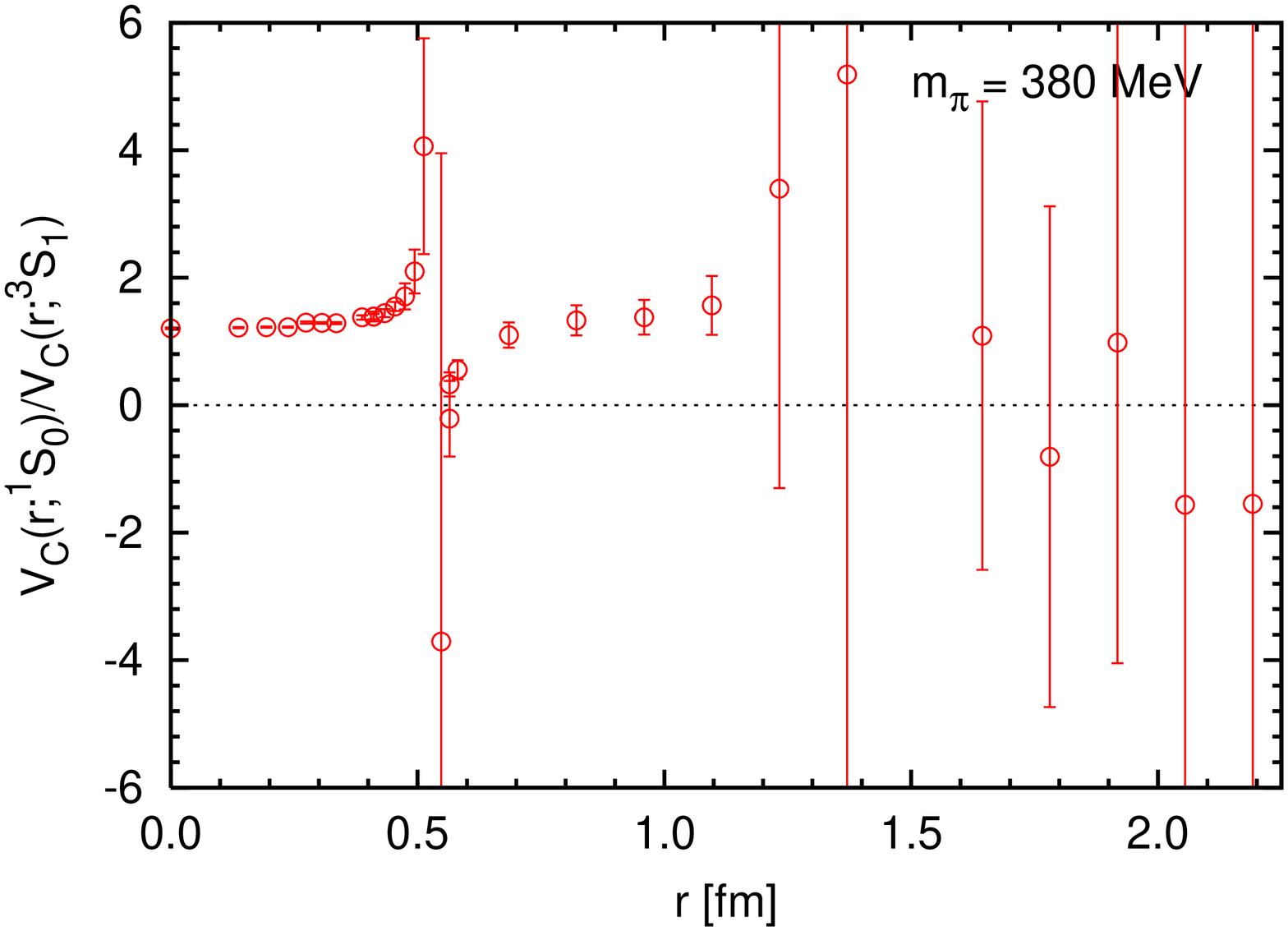}
\end{center}
\caption{
 The ratio  of the central potentials defined in
  Eq.~(\ref{eq:ratio-F})  for the lightest quark mass, $m_{\pi}=380$ MeV.}
\label{fig7}
\end{figure}
 
To check if there is an evidence of the 
 exponential tail from the dipole ghost in the long range part
  of the effective central potentials, the ratio
 ${\cal R}_{\rm 13}$ given by  Eq.~(\ref{eq:ratio-F})  is plotted 
 in Fig.~\ref{fig7} as a 
 function of $r$ for the lightest quark mass, $m_{\pi}=380$ MeV.
 Within the statistical errors, there is no sign that 
 ${\cal R}_{\rm 13}  \rightarrow -3$  for $r > 1$ fm,
 { so that possible ghost contamination  is small in our results
 with relatively heavy quark masses.}
  The figure also shows that ${\cal R}_{\rm 13} $ is rather close to $+1$ for
   $r > 0.7 $ fm.  This does not necessary implies that the 
   OPEP is seen: as long as there are spin-isospin independent 
    attraction such as originating from the 
    two-pion-exchange potential, it also leads to ${\cal R}_{\rm 13}  \simeq 1$.

\subsection{NN scattering lengths}

\begin{table}[t]
\caption{Effective center of mass energies $E=k^2/m_N$ obtained from the 
 asymptotic momenta for different  quark mass. $a_0$'s are the  
 associated scattering lengths obtained from L\"uscher's formula 
 Eq.~(\ref{eq:SL}). }
\begin{center}
\begin{tabular}{lllll}
\hline\hline
  $m_{\pi}$ [MeV]  & $E(^1{\rm S}_0)$ [MeV] & $E(^3{\rm S}_1)$ [MeV]
& $a_0(^1{\rm S}_0)$ [fm] & $a_0(^3{\rm S}_1)$ [fm] \\
\hline
 731.1(4) & $-0.400(83)$  & $-0.480(97)$ & $\ \ 0.115(26)$ & $\ \ 0.141(31)$\\
 529.0(4) & $-0.509(94)$  & $-0.560(114)$& $\ \ 0.126(25)$  & $\ \ 0.140(31)$\\
 379.7(9) & $-0.675(264)$ & $-0.968(374)$& $\ \ 0.153(66)$  & $\ \ 0.230(101)$\\
\hline
\end{tabular}
\end{center}
\label{table_a0}
\vspace{0.5cm}
\end{table}

As we found in Fig.~\ref{fig6}, the central potential multiplied by
 $r^2$ shows a net attraction as a result of the
 large cancellation between the short range 
 repulsion and the medium range attraction.
 This attractive nature of the  potential can be 
quantified by the scattering length $a_0$ defined from 
 L\"{u}scher's formula, Eq.~(\ref{eq:SL}), together with the 
 asymptotic momentum $k$ obtained from Eq.~(\ref{eq:GF}).\footnote{If
 the net interaction is small in the infinite volume limit, 
 the volume integral of the potential and the 
 scattering length are related in the Born approximation as, 
 $ a_0^\textrm{weak-coupling}\linebreak  \simeq  -m_N\int V_C(r) r^2 dr$.}

\begin{figure}[t]
\begin{center}
\includegraphics[width=7cm,angle=-90]{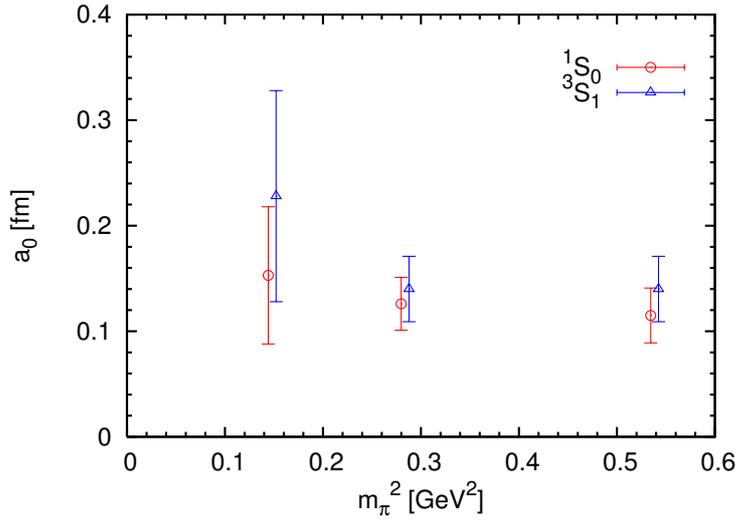}
\end{center}
\caption{
 Scattering length $a_0$ in the $^1{\rm S}_0$ and $^3{\rm S}_1$ channels for 
  three different quark masses obtained in the quenched QCD simulations.}
\label{fig9}
\end{figure}

The results of $a_0$ are summarized in the last two columns in 
Table \ref{table_a0} where $O(k^2)$ correction on the 
 right-hand side of Eq.~(\ref{eq:SL}) is assumed to
 be small for the present energy $E=k^2/m_N$.
 In Fig.~\ref{fig9}, the scattering lengths 
 for $^1 {\rm S}_0$ and $^3 {\rm S}_1$ channels are  shown as a function of $m_{\pi}^2$.
 Although there is a small attraction which increases as $m_{\pi}$ decreases
  in both channels, the absolute magnitudes of $a_0$ are much  smaller than
  the experimental values at the physical point:
   $a_0^{\rm(exp)}(^1{\rm S}_0)$ $\sim$  $20$ fm and $a_0^{\rm(exp)}(^3{\rm S}_1)\sim -5$ fm
    at $m_{\pi}^2 = 0.018$ GeV$^2$.
 
  The above discrepancy is partly 
  attributed to the heavy quark masses employed in our simulations:
   If we  can get closer  to the physical  quark mass in full QCD simulations,
   there should  arise the
 ``unitary region" where the  $NN$ scattering length becomes singular 
  and    changes    sign. This  
     was first noted in clear terms  by Kuramashi \cite{Kuramashi:1995sc} 
   and   was later elaborated in Refs.~\citen{NPLQCD} 
   and \citen{Epelbaum:2005pn} 
     by using chiral perturbation theory.     
     The singularity is  associated with  the  formation of the 
     di-nucleon bound state,
      so that the $NN$ scattering length becomes a  non-linear function
   of the quark mass in the unitary region.
    As suggested in Ref.~\citen{Kuramashi:1995sc} by using 
   the one-boson-exchange model  with 
   the quark-mass dependence of the hadron masses taken from the lattice QCD data,
   the size of the unitary region could be narrow, which implies that
   the scattering lengths at the heavy quark masses adopted in our simulation 
   can be as small as the values in Fig.~\ref{fig9}.
 
   Unlike the scattering length, the $NN$ potential would not have 
    singular behavior in the unitary region as expected from the well-known  
     quantum mechanical examples such as the low-energy scattering between 
     ultracold atoms. Also, the effective range parameter would be a rather
      smooth function  of the quark mass.  To check these points
    in QCD,  it is important to study the $NN$ potential, the scattering length
    and the effective range simultaneously in the full QCD simulations
{which allow us to approach small quark masses without quenched artifact.}
     Studies along this direction is now underway
      \cite{Ishii:2009zr} and 
     will be reported elsewhere.

\subsection{BS wave function in the ${\rm D}$-state}

\begin{figure}[t]
\begin{center}
\includegraphics[height=0.48\textwidth,angle=-90]{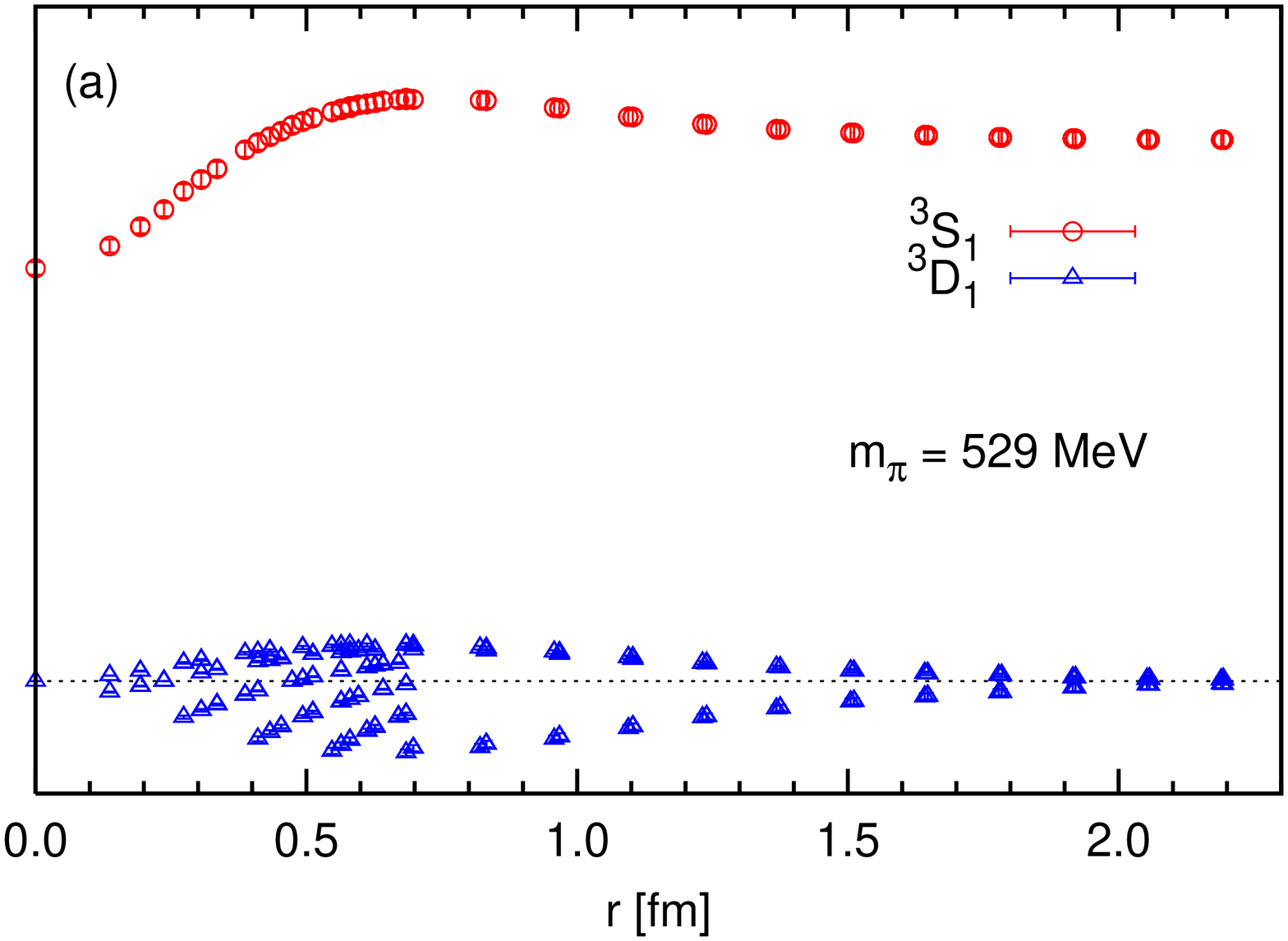}
\includegraphics[height=0.48\textwidth,angle=-90]{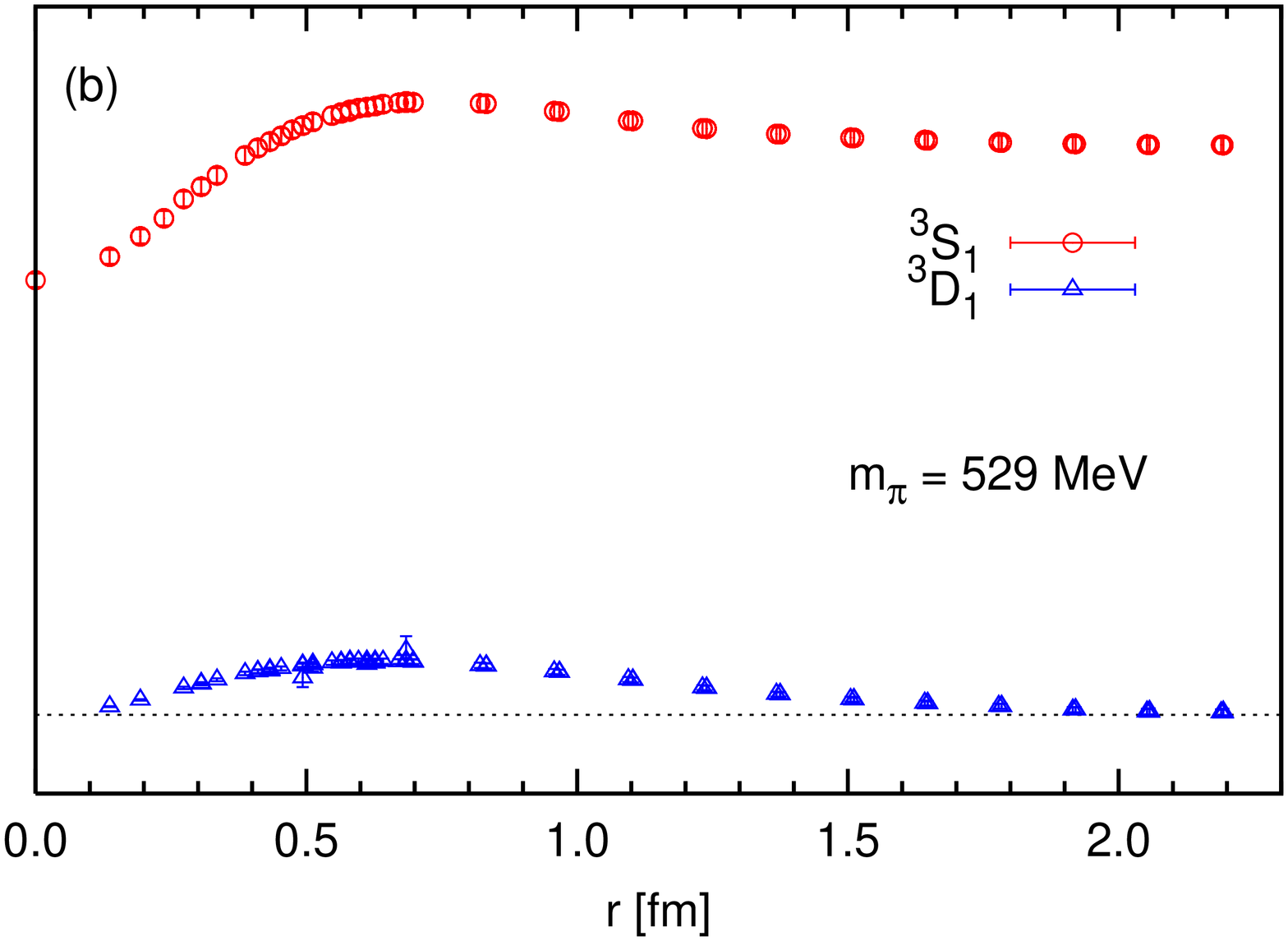}
\end{center}
\caption{(a) $(\alpha,\beta)=(2,1)$  components of the  S-state  and the  D-state
 BS wave functions projected out from a single state with $J^P=1^+, M=0$.
(b) The same data with the spherical harmonics components
 are removed in the D-state. }
\label{fig:tensor.wave}
\end{figure}

In \Fig{fig:tensor.wave}(a), we  show the $^3{\rm S}_1$ and $^3{\rm D}_1$ components
 of the  BS  wave functions obtained from the 
$J^P=1^+,  J_z=M=0$ state  for  $m_{\pi}\simeq 529$  MeV, according to the 
 procedure given in \S\ref{sec:tensor}.       
 To reduce  the computational cost,  the points are restricted  on the
coordinate  axes and  their nearest  neighbors  for $r  > 0.7$  fm,
whereas all points are calculated for $r < 0.7$ fm.

Note  that the $^3{\rm D}_1$  wave  function as a function of 
 $r$ is  multivalued  due to its angular dependence.
 Since  $(\alpha,\beta)=(2,1)$ spin
component of  the D-state wave function for  $J^P=1^+, M=0$ is
proportional to  the spherical harmonics
$Y_{20}(\theta,\phi)\propto   3\cos^2\theta  - 1$,   
it is a good consistency test to check if the multivaluedness can be absorbed by this
 angular  dependence.
Shown in \Fig{fig:tensor.wave}(b) are  the same BS wave functions
as \Fig{fig:tensor.wave}(a) with the angular dependence in the
D-state assumed to have this spherical harmonics form.
 It is clear that the multivaluedness is nicely removed, and thus
  it is certain that  we indeed extracted the D-state wave function
  on the lattice.

\subsection{Tensor force and its quark mass dependence} 

\begin{figure}[t]
\begin{center}
\includegraphics[width=8cm,angle=-90]{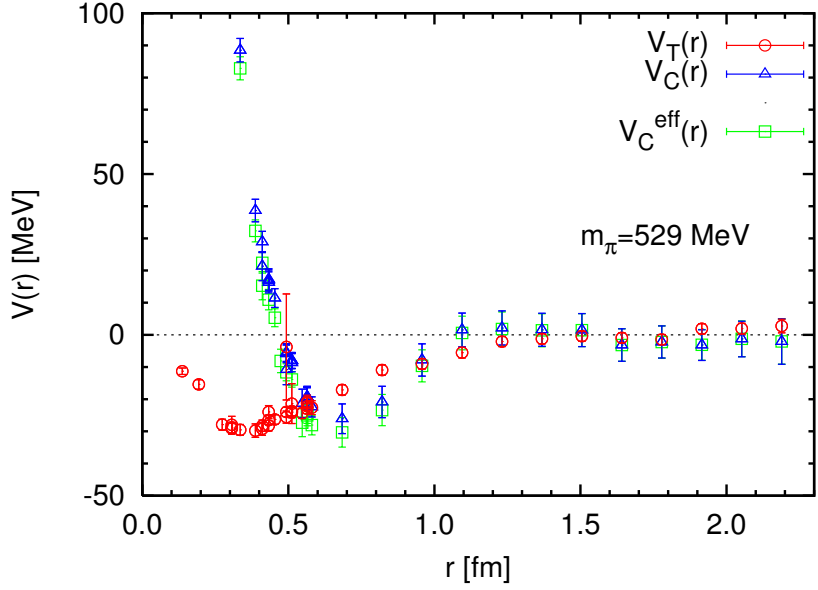}
\end{center}
\caption{The  central potential $V_C(r)$  and the tensor
potential 
$V_T(r)$ obtained from the $J^P=1^+$ BS wave function at
$m_{\pi}=529$ MeV. }
\label{fig:tensor.force}
\end{figure}

 Shown in \Fig{fig:tensor.force}  are the central
potential  $V_C(r)$ and  tensor potential  $V_T(r)$  together with
effective central potential $V_C^{\rm eff}(r)$ in the $^3 {\rm S}_1$ channel.
 (As mentioned before, we consider only the LO terms of the velocity expansion here
 by assuming  that the NLO term (the spin-orbit potential) and higher order terms
  are negligible at this low energy.)

Note  that $V_{\rm  C}^{\rm eff}(r)$  contains the  effect  of 
$V_T(r)$ implicitly as higher order effects through the process such as
 $^3{\rm S}_1 \rightarrow ^3{\rm D}_1 \rightarrow ^3{\rm S}_1$.
 In the real world,
 $V_{\rm  C}^{\rm eff}(r)$ is expected to acquire sufficient
attraction from the tensor force. This is the reason why
 bound deuteron exists in the $^3{\rm S}_1$ channel while the bound
  dineutron does not exist in the $^1{\rm S}_0$ channel. 
 Now, we see from \Fig{fig:tensor.force} that the difference between 
$V_C(r)$ and 
$V_C^{\rm eff}(r)$  is  still small in our quenched simulations due to 
 relatively large quark masses. This is also consistent with the 
 results of the small scattering length shown in \Fig{fig9}. 
%%%

\begin{figure}[t]
\begin{center}
\includegraphics[width=7.5cm,angle=-90]{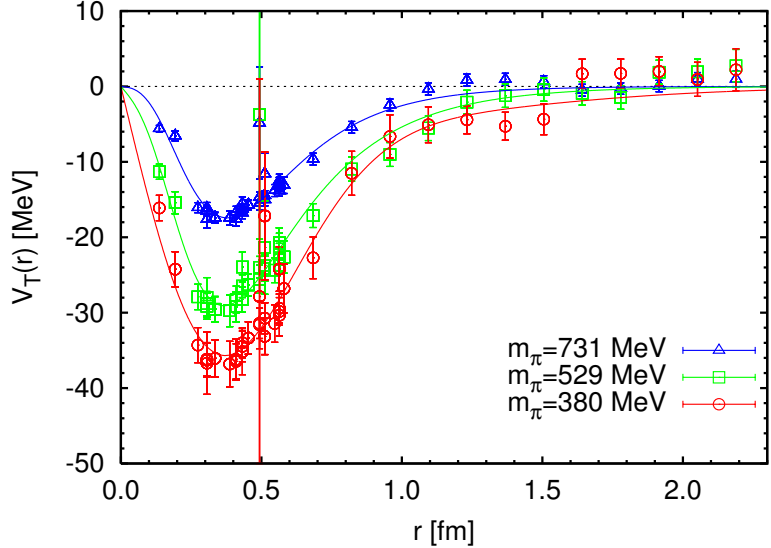}
\end{center}
\caption{Quark mass dependence of tensor force.
The lines are
 the four-parameter fit
  using the one-$\rho$-exchange +  one-pion-exchange with Gaussian form factors.}
\label{fig:tensor.force.mass}
\end{figure}

{  The  tensor potentials $V_T(r)$  in \Fig{fig:tensor.force}
  are negative for the whole range of $r$ within 
  statistical errors and have a minimum
 at short distance around  $0.4$ fm.}
%   If the long range part of the tensor force is dominated by 
{ If the tensor force receives significant contribution from}
the one-pion exchange as expected from the meson theory,
  $V_T(r)$ would be rather sensitive to the change of the
  quark mass.  As shown in \Fig{fig:tensor.force.mass},
   it is indeed the case:    Attraction of $V_T(r)$ is
    substantially enhanced as the quark mass decreases. 
 A phenomenological fit of the tensor force taking into account this physics
 will be given later.  
 
 As discussed in \S\ref{sec:ghost}, the ratio ${\cal R}_{\rm 13}$ of the 
  effective central potentials in the $^1{\rm S}_0$ and $^3{\rm S}_1$ channels
 is close to unity for $r > 0.7 $ fm so that we do not see 
  evidence of the dipole ghost (quenched artifact) in the 
 long range part of the potential with our relatively heavy quark masses.  
 However, this does not necessary imply that the OPEP
 is seen  in the effective central potentials: If the OPEP dominates at long distances,
   Eq.~(\ref{eq:OPEP-2}) immediately implies that the magnitude of the 
    tensor potential is always
   larger than the central potential at long distances. Since  
  this is not seen in Fig.~\ref{fig:tensor.force} within the
    statistical errors,  it is unlikely to interpret the 
     attraction of $V_C^{\rm eff}(r)$ at $0.5\ {\rm fm} < r < 1\ {\rm  fm}$
   as the evidence of OPEP.   
  
A technical comment is in order here.  Since  we use the $(\alpha,\beta)=(2,1)$
spin component of  \Eq{eq.vc},  the  second equation vanishes at 
 $\br \propto  (\pm 1,\pm  1,\pm  1)$.  
  This is  because  the spin  $(2,1)$
component of  the D-state  wave function   is
proportional         to      
$Y_{20}(\theta,\phi)\propto   3\cos^2\theta   -   1$  which
vanishes at $\br \propto (\pm 1,\pm 1,\pm 1)$.
Although these points are removed from our plots, statistical error is
accumulated in  the neighborhood of these points.   (For instance, see
the  points at  $r\simeq 0.5$  fm in  Figs.~\ref{fig:tensor.force} and
\ref{fig:tensor.force.mass}.)  A resolution of this problem by 
combining the data with other spin components will be reported
in the future publication.

 The central and tensor potentials obtained from lattice QCD are given
 at discrete data points.  For practical 
 applications  to nuclear physics,
 it is more useful to parametrize the lattice results 
 by known functions. We have tried such a fit for 
 $V_T(r)$ under the assumption of the one-$\rho$-exchange +
 one-pion-exchange with Gaussian form factors: 
\beq
V_T(r) &=& b_1 (1- e^{-b_2 r^2})^2
\left( 1 + \frac{3}{m_{\rho}r} + \frac{3}{(m_{\rho}r)^2} \right)  \frac{e^{-m_{\rho}r} }{r} 
\nonumber\\
& & +  b_3 (1- e^{-b_4 r^2})^2 \left( 1 + \frac{3}{m_{\pi}r} + \frac{3}{(m_{\pi}r)^2} \right)
  \frac{e^{-m_{\pi}r} }{r} ,
\label{eq:VT-par}
\eeq
where, $b_{1,2,3,4}$ are the fitting parameters while $m_{\rho}$ ($m_{\pi}$) is
 taken to be the 
$\rho$-meson mass (the pion mass)  calculated for each quark mass. 
  At this moment, it is hasty to extract physical quantities from the fit 
  such as the  meson-nucleon coupling constants: Nevertheless, it may be worth mentioning 
  that
  the pion-nucleon coupling constant extracted from the parameter $b_3$ in the 
   case of the lightest pion mass ($m_{\pi}=380$ MeV) reads
    $g_{\pi N}^2/(4 \pi) = 12.1 \pm 2.7$ which is encouragingly close to the 
    empirical value. We have tried similar fits for the central potential
     with the phenomenological repulsive core with a Gaussian form and the 
     meson-exchange potential with form-factors: The results are still
     not stable enough due to the statistical errors of the lattice data.

\subsection{Velocity dependence of the potential}
\label{sec:vel-dep}

  So far we have considered the potential determined from the 
  lattice data taken almost at zero effective 
    energy $E \simeq 0$ MeV (see  Table \ref{table_a0}).
  If the local potential  determined from the other energies has
  different spatial structure, it is an indication that there are
  velocity dependent terms as discussed in \S\ref{sec:2-body}.
  
A lattice  QCD analysis on  the velocity dependence has  been recently
carried out  by changing the  spatial boundary condition of  the quark
field  from the periodic  one to  the anti-periodic  one, so  that the
{  effective center  of mass energy  is increased to  $E \sim
 3 (\pi/L)^2/m_N \sim 50  $ MeV \cite{murano}.}
The result  shows that the central  and tensor potentials  do not show
modifications for every $r$ within the statistical errors: Namely, the
non-locality  of the potential  with our  choice of  the interpolating
operator is small and the potentials shown in the present paper can be
used  in the energy  region at  least up  to $E  \sim 50$  MeV without
significant modifications.\footnote{
An investigation  based on integrable models  suggests that potentials
derived  from the  BS wave  functions  with local  operators in  these
models are slowly varying functions of energy (velocity) \cite{ope}. }
Detailed  account of  the above  result is  beyond the  scope  of this
paper, and will be reported elsewhere.
%

%%%%%%%%%%%%%%%%%%%%%%%%%%%%%%%%%%%%%%%%%%%%
\section{Summary and concluding remarks}
\label{sec:summary}

 In this paper, we have discussed  
  the basic notion of the nucleon-nucleon potential and its field-theoretical 
  derivation from the   equal-time Bethe-Salpeter wave function in QCD. 
  By construction, the non-local potential defined through the projection of the
   wave function to the interaction region (the inner region)   correctly reproduces the 
   asymptotic form of the wave function in the region beyond the 
   range of the nuclear force (the outer region). Thus the observables such as the 
   phase shifts and the binding energies can be
   calculated after extrapolating the potential to the infinite volume limit.
   Non-locality of the potential can be taken into account successively by
    making its velocity expansion, which introduces the velocity-dependent
     local potentials. The leading-order terms  of such velocity expansion 
      for the 
      nucleon-nucleon interaction are the central and the tensor potentials.
           
  As an exploratory study to test how this formulation works, we have
  carried out  quenched lattice  QCD simulations of the two-nucleon system
  in a spatial box of the size (4.4 fm)$^3$ with the quark masses 
  corresponding to $m_{\pi} = 380, 529, 731$ MeV. 
   We  found that the $NN$ potential calculated on the lattice at low energy 
  shows all the characteristic features 
  expected from the empirical $NN$ potentials obtained from the experimental
   $NN$ phase shifts, namely  the 
  attractive well  at  long and medium distances and the repulsive core 
  at short distance for the 
   central potential. As for the tensor potential obtained from the  
  coupled channel treatment of the $^3{\rm S}_1$-state
  and the $^3{\rm D}_1$-state in the BS wave functions on the lattice,
 we found appreciable attraction at long and medium distances and 
 a  moderate repulsion at short distance.  
 
 As the quark mass decreases, the repulsive core and 
   attractive well in the central potential, and the attractive well in the 
   tensor  potential tend to be enhanced.  Also, we found net attraction
   in both $^1{\rm S}_0$ and $^3{\rm S}_1$ channels after the cancellation of the 
    repulsive core and the attractive well.  The absolute magnitudes
      of the scattering lengths are still much smaller than the 
       physical values due to the large quark mass in our simulation.
       Phenomenological fit of the tensor potential
       strongly suggests the existence of the one-pion-exchange
       contribution in its long range part.

  There are a number of  directions to be investigated
   on the basis of our approach as listed below:
\begin{itemize}
  \item[1.] Determination of the velocity dependence  is
   important in deriving the $NN$ potentials which can be used for the wide
    range of scattering energies.
  Studies along this line using the anti-periodic boundary condition in the spatial direction
  has been already started  \cite{murano} as mentioned in \S\ref{sec:vel-dep}.
  \item[2.] To derive the realistic  $NN$ potentials on the lattice,
    it is necessary to 
    carry out full QCD simulations with  dynamical quarks.
   Studies along this line  with the use of the  
    (2+1)-flavor QCD configurations with the Wilson fermion  generated by PACS-CS
    Collaboration \cite{Kuramashi:2008tb}
 is currently under way \cite{Ishii:2009zr}.
  \item[3.] The hyperon-nucleon ($YN$) and hyperon-hyperon ($YY$)
   potentials are essential for understanding the properties
    of hyper nuclei and the hyperonic matter inside the
     neutron stars.  However, the experimental scattering data are
      very limited due to the short life-time of hyperons.  
 On the other hand, the $NN$, $YN$ and $YY$ interactions on the lattice 
  can be treated in the same manner by changing only the quark flavors.
   Recently, 
  the $\Xi N$ potential in quenched QCD \cite{Nemura:2008sp}
   and the $\Lambda N$ potential in quenched and full QCD \cite{Nemura:2009kc}
 are examined   as a first step toward systematic derivation of the hyperon
   potentials.
  \item[4.] The three-nucleon force is thought to play important roles in
 nuclear structures and  in the equation of state 
 of  high density matter as mentioned in \S \ref{sec:many-body}. 
  Since the experimental information is
 scarce, simulations of the three nucleons  on the lattice 
 combined with the method proposed in \S \ref{sec:many-body} may
 lead to the first principle determination of the three-nucleon 
 potential in the near future.   
\end{itemize}

    If it turns out that the program described in this paper indeed works in full QCD
     with realistic quark masses, it would be 
      the promising first step toward the understanding 
   of atomic nuclei and neutron stars   from the fundamental law of the strong
   interaction, the quantum chromodynamics.

%%%%%%%%%%%%%%%%%%%%%%%%%%%%%%%%%%%%%%%%%%%%%%%%%%%%%%%%%%%%%%%%%%%%%%%%%%%%%%%

\section*{Acknowledgements}

The authors thank T. Doi, E. Hiyama, T. Inoue, Y. Ikeda, N. Ishizuka, A. Jackson, 
K. Murano, H. Nemura, S. Nishizaki, 
M. Oka, T. Otsuka,  K. Sasaki, S. Sasaki, T. Takatsuka,
R. Tamagaki, K. Yabana, Y. Yamamoto and W. Weise
for useful discussions and comments.
This  research was  supported  in  part
by  the  Grant-in-Aid  of  MEXT (Nos.~15540254,  18540253,  19540261,
20340047)
and by a Grant-in-Aid for Specially Promoted Research (No.~13002001)
and  by  Grant-in-Aid  for  Scientific Research  on  Innovative  Areas
(No.~2004: 20105001,20105003).
  Our simulations have been performed with IBM Blue
Gene/L at KEK under a support of its Large Scale Simulation Program, No.~18 and
No.~06-21 (FY2006), No.~07-07 (FY2007), No.~08-19 (FY2008) and  No.~09-23 (FY2009).

\appendix

%%%%%%%%%%%%%%%%%%%%%%%%%%%%%%%%%%%%%%%%%%%%%%%%%%%%%%%%%%%%%%%%%%%%%%%%%%%%%%%
\section{Bethe-Salpeter Wave Function and Its Asymptotic Behaviour}
\label{sec:BS-infinite}
%%%%%%%%%%%%%%%%%%%%%%%%%%%%%%%%%%%%%%%%%%%%%%%%%%%%%%%%%%%%%%%%%%%%%%%% 
In this appendix we derive the behaviour of the Bethe-Salpeter (BS) wave function at large $r$, only using the properties of quantum field theories.

\subsection{Unitarity of $S$-matrix and structure of $T$-matrix}
\label{sec:A-1}
We first determine the structure of the $NN$ scattering $T$-matrix below the
 pion-production threshold. Due to the unitarity  of the $S$-matrix.
%\begin{eqnarray}
$S^\dagger S = 1$ with $ S= 1+  i T$,
%\end{eqnarray}
we obtain
\begin{eqnarray}
\langle f\vert T \vert i \rangle - \langle f\vert T^\dagger \vert i \rangle&=& 
i \sum_n \langle f \vert T^\dagger  \vert n \rangle \langle n \vert T \vert i \rangle .
\label{eq:unitarity}
\end{eqnarray}
In the case of $NN$ scattering in the center of mass frame
such that $(k_a,s_a) + (k_b,s_b) \rightarrow (k_c,s_c)+(k_d,s_d)$
where $k_a=(\varepsilon_k,\bk)$, $k_b=(\varepsilon_k,-\bk)$ 
and $k_c=(\varepsilon_p,\bp)$, $k_d=(\varepsilon_p,-\bp)$
with $\varepsilon_k =  \sqrt{\bk^2+m_N^2}$ and $\varepsilon_p= \sqrt{\bp^2+m_N^2}$,
we write
\begin{eqnarray}
\! \! \! \! \! \! \! \!
 _{\rm in}\langle  p_c, s_c,  p_d,s_d\vert T \vert  p_a, s_a, p_b, s_b\rangle_{\rm in} &=&
(2\pi)^4 \delta^{(4)}(p_a+p_b-p_c-p_d)T(\bp,s_c,s_d;\bk,
s_a,s_b).\nonumber \\
\end{eqnarray}
Here $s_i=\pm 1/2$ is a helicity of each nucleon, and $k=\vert \bk\vert = \vert \bp\vert$ in the center of mass frame. Below the pion production threshold such that 
$ 2\sqrt{k^2+m_N^2} < 2 m_N + m_\pi$, the sum over intermediate states $n$ in  
Eq.~(\ref{eq:unitarity}) can be  restricted to  the $NN$ states due to
 energy-momentum conservations as
\begin{eqnarray}
\sum_n \vert n \rangle \langle n\vert &=&
\sum_{s_1,s_2} \int \frac{d^3 p_1}{(2\pi)^3 2\varepsilon_{p_1}}
\frac{d^3 p_2}{(2\pi)^3 2\varepsilon_{p_2}}
\vert p_1,s_1,p_2, s_2  \rangle \langle p_1,s_1,p_2, s_2 \vert .
\end{eqnarray}
This leads to
\begin{eqnarray}
&&\hspace{-7mm}T(\bp,s_c,s_d;\bk, s_a,s_b) - T^\dagger(\bp,s_c,s_d;\bk, s_a,s_b) 
\nonumber \\
&&=  i\sum_{s_1,s_2} \frac{k}{32\pi^2 \varepsilon_k} 
\int d\Omega_{q}\ T^\dagger(\bp,s_c,s_d;\bq, s_1,s_2)
T(\bq,s_1,s_2;\bk, s_a,s_b) ,
\end{eqnarray}
where $\vert\bq\vert = k$ and $\Omega_{q}$ is the solid angle of vector $\bq$.
Using the angular momentum basis\cite{JW},
\begin{eqnarray}
T(\bp,s_c,s_d;\bk, s_a,s_b) &=&  4 \pi \sum_{J,M}  \frac{2J+1}{4\pi} 
 \langle s_c,s_d\vert T^J(k)
\vert s_a,s_b\rangle
(D^J)_{ s^\prime  M}^\dagger(\Omega_{p}) D_{ M s }^J(\Omega_{k}) ,\nonumber\\
\end{eqnarray}
with $s=s_a-s_b$ and $s^\prime = s_c-s_d$,
we obtain
\begin{eqnarray}
T^J(k) - [T^{J}]^\dagger (k) =  i \frac{k}{8\pi \varepsilon_k} [T^J]^\dagger (k) T^J(k).
\end{eqnarray}
Here $T^J$ is considered as a $4\times 4$ matrix and 
the Wigner $D$-matrix $D^J$ is defined by
\begin{eqnarray}
D^J_{M \lambda}(\Omega) &=&  e^{- iM \alpha} d^J_{M \lambda }(\beta) e^{+ i\lambda  \alpha} ,
\end{eqnarray}
where the solid angle is denoted as $d\Omega =\sin\beta d\beta d\alpha$ 
 and $d^J_{M \lambda}(\beta)$ is the Wigner $d$-matrix.
  The normalization of the $D$-matrix is given by
\begin{eqnarray}
\int d\,\Omega \ (D^{J})^\dagger_{\lambda M}(\Omega)D^{J'}_{M' \lambda}(\Omega)
&=& \frac{4\pi}{2J+1}\delta^{JJ'}\delta_{M M'} ,
\label{eq:orthogonal}
\end{eqnarray}
where no summation is taken for $\lambda$.
For the $NN$ scattering, with new helicity basis such that 
$\vert +\frac{1}{2},+\frac{1}{2} \rangle \pm \vert -\frac{1}{2},-\frac{1}{2} \rangle$ 
and $\vert +\frac{1}{2},-\frac{1}{2} \rangle \pm \vert -\frac{1}{2},+\frac{1}{2}\rangle$, 
$T^J$ is decomposed into two $1\times 1$ submatrices and 
and one $2\times 2$ submatrix as \cite{JW}
\begin{eqnarray}
T^J &=& \left(
\begin{array}{ccc}
T_{\ell=J, s=0}^J & 0 & 0_{1\times 2} \\
0 & T_{\ell=J, s=1}^J & 0_{1\times 2} \\
0_{2\times 1} & 0_{2\times 1} & T_{\ell=J\mp 1, s=1}^J \\
\end{array}
\right).
\end{eqnarray}
The unitarity condition then gives
\begin{eqnarray}
T^J_{\ell=J,s} = \hat T_{J s} , \ \ \ 
T_{\ell=J\mp 1, s=1}^J  =  O(k) \left(
\begin{array}{cc}
\hat T_{J-1,1} & 0 \\
0 & \hat T_{J+1,1} \\
\end{array}
\right) O^{-1}(k) ,
\end{eqnarray}
with
\begin{eqnarray}
\label{eq:T-form}
\hat T_{\ell s}  = \frac{16\pi \varepsilon_k}{k}
e^{i\delta_{\ell s}(k)}\sin \delta_{\ell s}(k) ,
 \ \ \ 
O(k) = \left(
\begin{array}{cc}
\cos \varepsilon_J(k) & -\sin\varepsilon_J(k) \\
\sin\varepsilon_J(k) & \cos\varepsilon_J(k) \\
\end{array}
\right) , 
\end{eqnarray}
where $\delta_{\ell s}(k)$ is the scattering phase shift, whereas
 $\varepsilon_J(k)$ is  the mixing angle between  $\ell=J\pm 1$.  They 
correspond to the standard Blatt-Biedenharn eigenphase and mixing angle.

\subsection{BS amplitude and half off-shell $T$-matrix}
Let  us now consider the Bethe-Salpeter (BS) amplitude for the proton and the neutron, 
defined by
\begin{eqnarray}
\Psi_{\alpha \beta}(x,y)  
= \langle 0 \vert {\rm T}\{ n_{\beta}(y) p_{\alpha}(x) \}\vert {\rm p}(\bq, s) {\rm n}(\bq', s') \rangle_{\rm in} ,
\label{eq:BS-def}
\end{eqnarray}   
where   ${\rm T}$ represents the time-ordered product.
The spatial momentum and the helicity for the incoming proton and those for 
 the neutron are
  denoted by  $(\bq, s)$ and $(\bq', s')$, respectively.  The single nucleon
   state is normalized covariantly,\linebreak 
  $\langle B_i(\bq, s)\vert B_j(\bq', s' )\rangle 
  =  2\varepsilon_{q} (2\pi)^3\delta_{ij}\delta_{ss'} \delta^3(\bq-\bq')$ 
  where $B_1=p$ (proton) and $B_2=n$(neutron).
 
The fields, ${n}_{\beta}(y)$ and ${p}_{\alpha}(x)$, are the local composite operators for the 
  neutron and the proton whose explicit forms are irrelevant for the following derivation. 
  One of the advantages to use 
  local operators is that the standard reduction formula can be generalized 
  without much modification as shown by  Nishijima, Zimmermann and Haag (NZH) \cite{NZ}.
  In particular, one can define  in and out composite fields, ${n}_{\rm in (out)}(x)$
   and ${p}_{\rm in (out)}(x)$,
  in a similar way as the elementary field through the Yang-Feldman equation as \cite{NZ}
 \begin{eqnarray}
\sqrt{Z} N_{\rm in (out)}(x) = N(x) - \int S_{\rm ret (adv)}(x-x';m) J (x') d^4 x ,
 \end{eqnarray} 
where $N$ takes either  ${n}$ or ${p}$, $S_{\rm ret (adv)} $ denotes the 
retarded (advanced) Green's function in the free space 
with the mass $m=m_N$,   and the ``source" is $J(x) \equiv (i \dslash_x - m) N(x)$.
The wave function renormalization constant $Z$ is defined as 
$\sqrt{Z} u_{\alpha}(\bp,s)= \langle 0 | N_{\alpha}(0)|B(\bp,s) \rangle$,
 where we have the following normalization of the Dirac spinors:
\beq
\sum_{\alpha} u_{\alpha}^{\dagger}(\bp,s)u_{\alpha}(\bp,s')
& = & \sum_{\alpha} v_{\alpha}^{\dagger}(\bp,s)v_{\alpha}(\bp,s') 
= 2\varepsilon_{p} \delta_{ss'}, \\
\sum_{s} u_{\alpha}(\bp,s)\bar{u}_{\beta}(\bp,s)
& = & (\pslash + m)_{\alpha \beta}, \ \ \ 
\sum_{s} v_{\alpha}(\bp,s) \bar{v}_{\beta}(\bp,s)
 = (\pslash - m)_{\alpha \beta}. 
\eeq
 
Then the  NZH reduction formula is summarized as
 \begin{eqnarray}
& &\hspace{-7mm}
  \sqrt{Z} \left[ {\rm T}({\cal O}) B_{\rm in}^{\dagger}(\bp,s)
  - (-)^{|{\cal O}|} B_{\rm out}^{\dagger}(\bp,s) {\rm T}({\cal O}) \right] 
 \nonumber \\
& &  =  \int d^4 x\, e^{-ipx}\ {\rm T}\{{\cal O} \bar{N}(x)\} [-i S^{-1}(p) u(\bp,s)] , \\
&&\hspace{-7mm}
\sqrt{Z} \left[ B_{\rm out}(\bp,s) {\rm T}({\cal O}) 
  - (-)^{|{\cal O}|}  {\rm T}({\cal O}) B_{\rm in}(\bp,s)\right]
 \nonumber \\
& &  =  \int d^4 x\, e^{ipx}\  [-i \bar u(\bp,s) S^{-1}(p) ] {\rm T}\{N(x){\cal O}\} .
\label{eq:HNZ}
\end{eqnarray}
Here ${\cal O}$ is an arbitrary  product of operators with 
 the number of fermionic operators denoted by $ |{\cal O}|$, and 
 $S^{-1}(p)=(\pslash-m+i\delta)$ is  the inverse of the free nucleon propagator.
 The asymptotic baryon and anti-baryon operators,
   $B_{\rm as}(\bp,s)$ and $D_{\rm as}(\bp,s)$  (${\rm as} = {\rm in, out}$)
   are defined by the Fourier decomposition of $N_{\rm as}(x)$,
 \begin{eqnarray}
 \label{eq:B_in}
 N_{\rm as} (x) = \sum_s 
   \int \frac{d^3 p}{{(2\pi)^3 2\varepsilon_{p}}} 
   \left[ e^{-ipx}\   B_{\rm as}(\bp,s) u(\bp,s) 
     + e^{ipx}\ D_{\rm as}^{\dagger}(\bp,s) v(\bp,s) \right] ,\qquad
 \end{eqnarray}
where the flavor and spinor indices are suppressed.
The operator $B_{\rm as}$ thus defined satisfies the covariant commutation relation,
$\{ B_{\rm as }(\bp,s), B_{\rm as}^{\dagger}(\bp',s') \} = 2\varepsilon_{p} (2\pi)^3
\delta_{ss'} \delta^3(\bp-\bp')$, and
asymptotic states are defined by
$\vert B(\bp,s ) \rangle_{\rm as} =  B^\dagger_{\rm as}(\bp,s )\vert 0 \rangle $.

By using the NZH reduction formula,
we can evaluate our BS amplitude Eq.~(\ref{eq:BS-def}) as
\begin{eqnarray}
&&\Psi_{12}(x_1,x_2) =\nonumber\\
&& Z^{-1} \int \prod_{i=1}^2 \left\{ \frac{d^4 q_i}{(2\pi)^4}   e^{ -i q_ix_i} \right\}\
 G_{12;34}  (q_1,q_2;q_3,q_4)
[-iS^{-1}(q_3) u(3)]_3[ -i S^{-1}(q_4) u(4)]_4 , \nonumber \\
\label{eq:A2-BS0}
\end{eqnarray}
where   the four-point Green's function is defined by   
\begin{eqnarray}
\label{eq:G4}
& & G_{12;34}(q_1,q_2; q_3,q_4)= \int \prod_{i=1}^4 \left\{d^4x_i
e^{ i q_ix_i} \right\}\
\langle 0 \vert {\rm T}\{ n_2(x_2) p_1(x_1) \bar{p}_3(x_3) \bar{n}_4
(x_4)\} \vert 0 \rangle .\nonumber \\
 \end{eqnarray}  
Here, to simplify the notation, we abbreviate the 
Lorentz indices by  the lower-case suffixes  ($1, \cdots, 4$) with 
 the repeated suffixes being contracted
 and the state labels  are abbreviated as the numbers in the parenthesis, e.g.
 $u_{\alpha}(\bq,s) \rightarrow u_3(3)$ and $u_{\beta}(\bq',s') \rightarrow u_4(4)$. 
The four-point function can be decomposed into the 
 free part  and the 
 connected part as $G_{12;34} =Z^2 (G_{12;34}^{\rm (0)} + G_{12;34}^{({\rm c})} )$. 
  The free part reads
\beq 
 G_{12;34}^{\rm (0)}
 = (2\pi)^8 \delta^4(q_1-q_3) \delta^4(q_2-q_4) [iS(q_3)]_{13}[iS(q_4)]_{24},
 \eeq
whereas  the connected part is rewritten with the 
 proper vertex $\Gamma$ as
\beq
\! \! \! \! \! & & G_{12;34}^{\rm (c)} (q_1,q_2; q_3,q_4) \nonumber \\
\! \! \! \! \! & & =  (2\pi)^4 \delta^4(K-Q) 
\ [iS(q_1)]_{11'}
   [iS(q_2)]_{22'} \ (-i) \Gamma_{1'2';3'4'}(k; q|Q)  \ [iS(q_3)]_{3'3} [iS(q_4)]_{4'4} .\nonumber\\
\eeq
Here we have introduced  relative and center-of-mass (c.m.) 
4-momenta by
\beq 
K=q_1+q_2,\ \  k=(q_1-q_2)/2,\ \   Q=q_3+q_4,\ \  q=(q_3-q_4)/2.
\eeq

Then,  the $K$-integration in  Eq.~(\ref{eq:A2-BS0}) can be carried out to obtain
\beq
\label{eq:A2-BS1}
\! \! \! \! \! 
\Psi_{12}(x_1,x_2) &=& 
 \left[ \psi_{12}^{(0)}(r) + \psi_{12}^{\rm (c)}(r) \right] e^{-iQR} ,\\
\label{eq:A2-BS2}
\! \! \! \! \! 
\psi_{12}^{(0) }(r) & = &  Z u_1(3) u_2(4) e^{-iq r} , \\
\label{eq:A2-BS3}
\! \! \! \! \! 
\psi_{12}^{({\rm c})}(r) & = & i  Z
\int \frac{d^4 k}{(2\pi)^4} \ e^{-ik r} \ [S(q_1)]_{11'}[S(q_2)]_{22'}
 \Gamma_{1'2';34} (k; q|Q) u_3(3)u_4(4) ,\nonumber\\
 \eeq
where $r= x_1-x_2$ and $R=(x_1+x_2)/2$ are  relative and c.m. 4-dimensional coordinates, respectively.
 Covariant Nambu-Bethe-Salpeter type differential equation can be obtained
  by multiplying  $S^{-1}(i\dslash_1) S^{-1}(i\dslash_2)$ to 
 Eqs.~(\ref{eq:A2-BS1})$-$(\ref{eq:A2-BS3}) from the left:
\beq
\label{eq:BS-eq}
& & [S^{-1}(i\dslash_x)]_{\alpha \alpha'} [S^{-1}(i\dslash_y)]_{\beta \beta'}
  \Psi_{\alpha' \beta'}(x,y)\nonumber\\
&&    = i  Z \int \frac{d^4 k}{(2\pi)^4}  e^{-ik r} e^{-iQR}\
  \Gamma_{\alpha \beta;\gamma \delta}(k;q|Q)   u_\gamma(\bq,s)u_\delta(\bq',s'). \eeq

In our applications of the $NN$ scattering at low energies, 
it is useful to consider the equal-time BS amplitude (which we call the 
 BS wave function in the text) and associated
 Lippmann-Schwinger type integral equation or the   Schr\"{o}dinger type 
 differential equation.
  For this purpose, we first carry out  the  integration over $k^0$ in
 Eq.~(\ref{eq:A2-BS3}) using the explicit form of the free propagator:
\beq
\label{eq:S-prop}
\! \! \! \! \! \! \! 
S(p) = \left( \frac{1}{\pslash - m + i \delta}\right)_{\alpha \beta}
= \frac{1}{2\varepsilon_{p}} \left[
   \frac{\sum_s u_{\alpha}(\bp,s) \bar{u}_{\beta}(\bp,s)}{p^0 - \varepsilon_{p} + i \delta}
+  \frac{\sum_s v_{\alpha}(-\bp,s) \bar{v}_{\beta}(-\bp,s)}{p^0 + \varepsilon_{p} - i \delta} 
\right] .\nonumber\\
\eeq
 Since we are interested in the asymptotic form of the wave function at
  $|\br| \rightarrow \infty$ below
  pion production threshold, we can pick up only the nucleon pole from 
 $S(p)$  in the $k^0$-integral of Eq.~(\ref{eq:A2-BS3}) without loss of generality.
 Possible  poles from $\Gamma$ associated with 
  the resonance production and with the deuteron bound state,  as well as
   anti-nucleon poles in $S(p)$ in Eq.~(\ref{eq:S-prop}), 
  modify only the short-distant part of the wave function. 
 This does not at all imply that those contributions are not 
  important. They do affect the actual values of the phase shifts and mixing parameters
 and are fully taken into account in the definition of our potential, 
 Eqs.~(\ref{eq:QCD_nonlocal-potential}) and (\ref{eq:QCD_local-potential}).  
 
 Using the residue theorem and taking the equal-time limit ($x_0=y_0\equiv t$) in the 
  rest frame of the two-particles ($\bQ=0$), we end up with the 
  Lippmann-Schwinger type equation;    
\begin{eqnarray}
\label{eq:LS-1}
\! \! \! \! \! \! \! \! \! \! 
\Psi_{\alpha \beta}(\br,t) &=& \psi_{\alpha \beta}(\br; \bq,s,s')\ e^{-2i\varepsilon_{q}t} ,\\
\label{eq:LS-3}
\! \! \! \! \! \! \! \! \! \! 
\psi^{(0)}_{\alpha \beta}(\br; \bq,s,s') &=& Z u_{\alpha}(\bq,s) u_{\beta}(-\bq,s') 
e^{i\bqs \cdot \brs},
\\
\label{eq:LS-2}
\! \! \! \! \! \! \! \! \! \! 
\psi_{\alpha \beta}(\br; \bq,s,s') & = & \ \ 
\psi^{(0)}_{\alpha \beta}(\br; \bq,s,s')  \nonumber\\
&&+\sum_{\tilde{s},\tilde{s}'} \int \frac{d^3 k}{(2\pi)^3} 
 {\psi^{(0)}_{\alpha \beta}(\br; \bk,\tilde{s},\tilde{s}')} 
\frac{\varepsilon_q+\varepsilon_k}{8 \varepsilon_k^2}
\frac{ {\cal T}_{\tilde{s} \tilde{s}'; ss'}(\bk; \bq)} {\bk^2-\bq^2 - i\delta} 
+ {\cal I}(\br). %\nonumber \\
\eeq 
Here ${\cal I}(\br)$ originates from the contributions other than the nucleon pole
 and is an exponentially localized function in $\br$ below inelastic 
 threshold \cite{Lin:2001ek}.
 In Eq.~(\ref{eq:LS-2}), we have defined the  half off-shell $T$-matrix,
\begin{eqnarray}
 i{\cal T}_{12;34}(\bk;\bq) = \bar u_1(1) \bar{u}_2(2) (-i) \Gamma_{12;34}(k;q|Q) u_3(3)  u_4(4) , 
\end{eqnarray}
 where the outgoing energy $2\varepsilon_k=2\sqrt{ \bk^2+m^2}$ is not necessary equal to the  
  incoming energy $2\varepsilon_q=2\sqrt{ \bq^2+m^2}$.
 The   Schr\"{o}dinger type 
 differential equation is obtained from Eq.~(\ref{eq:LS-2}) by
   multiplying $\bq^2 + \nabla^2$,
 \beq
 \label{eq:KG-eq}
(\bq^2 + \nabla^2) \psi_{\alpha \beta}(\br; \bq,s,s') = -
 \sum_{\tilde{s},\tilde{s}'} \int \frac{d^3k}{(2\pi)^3 } 
 {\psi^{(0)}_{\alpha \beta}(\br; \bk,\tilde{s},\tilde{s}')}
\frac{\varepsilon_q+\varepsilon_k}{8 \varepsilon_k^2}  {\cal T}_{\tilde{s} \tilde{s}'; ss'}(\bk; \bq) + {\cal K}(\br),
\nonumber \\
 \eeq 
 with  ${\cal K}(\br)=(\bq^2 + \nabla^2){\cal I}(\br)$.
 Since the plain wave part of  $\psi_{\alpha \beta}(\br; \bq,s,s')$ is projected
  out by the operator $(\bq^2 + \nabla^2)$,  
   the right-hand side of Eq.~(\ref{eq:KG-eq}) is exponentially localized in
   $\br$ and vanishes for $r > R$. 
  
\subsection{Asymptotic BS wave function and the phase shift}
{ Let us further consider the asymptotic behaviour of  
$ \psi_{\alpha \beta}(\br; \bq,s,s') $ at large $r$ to relate it to  
the  scattering parameters (phase shifts and mixing angles) defined in \S\ref{sec:A-1}.
 The derivation of this subsection has been essentially
  given by Ishizuka in Ref.~\citen{ishizuka2}.}
  
To perform the $\bk$ integration in Eq.~(\ref{eq:KG-eq}), we 
introduce the following helicity decomposition of the half-off shell $T$-matrix,
 \begin{eqnarray}
\label{eq:TJ-kq}
&&{\cal T}_{s_1s_2; s_3s_4}(\bk,\bq) = 4\pi \sum_{J,M} \frac{2J+1}{4\pi}
 \langle s_1,s_2\vert T^J(k;q)
\vert s_3,s_4\rangle
(D^J)^\dagger_{ s  M}(\Omega_{k}) D_{ M s^\prime }^J(\Omega_{q}), \nonumber\\\\
\label{eq:UU}
&&u_{\alpha}(\bk,s_1) u_{\beta}(-\bk, s_2) e^{i\bks\cdot\brs} 
= \sum_{JM} D^{J}_{Ms}(\Omega_{k})
 U_{\alpha a}(\nabla) U_{\beta b}(-\nabla) \phi_{JMs_1s_2;ab}^{[j]}(\br, k)   ,
\nonumber \\ \end{eqnarray}
where $s=s_1-s_2$, $s^\prime =s_3-s_4$, $k=\vert\bk\vert $
 and $ q=\vert\bq\vert$.  
 The reduced wave function $\phi^{[j]}$ in the $2\times 2$ spinor space
  labeled by the indices $a,b$ is defined as
\begin{eqnarray}
\label{eq:JMLL}
\phi_{JM\lambda_1\lambda_2}^{[j]}(\br, k) &=&\sum_{\ell , s} \phi_{JM \ell s}^{[j]}
 (\br, k)\langle JM \ell s\vert JM \lambda_1\lambda_2\rangle  ,  \\
\label{eq:chi-ab}
\phi_{JM \ell s}^{[j]}(\br, k) &=& j_\ell (kr) Y_{JM}^{\ell s}(\Omega_{r}),
\
Y_{JM}^{\ell s}(\Omega_{r}) 
=\sum_{\ell_z s_z} Y_{\ell \ell_z}(\Omega_{r}) \chi(s,s_z)
\langle \ell s \ell_z s_z\vert JM\rangle .\nonumber\\
\end{eqnarray}
Note that  $U_{\alpha a}(\nabla)$ and $U_{\beta b}(-\nabla)$ in Eq.~(\ref{eq:UU})  are
the  $4 \times 2$ matrices acting on the $2 \times 2$ matrix 
$\phi_{ab}^{[j]}$ so that the Dirac structure $u_{\alpha}(\bk,s_1) u_{\beta}(-\bk,s_2)$
 is correctly reproduced:  
  Explicitly, 
  $U({\bf \nabla}) 
= \sqrt{(\varepsilon_k+ m_N)} ( I_{2\times 2}, -i \bsigma \cdot \nabla/(\varepsilon_k+ m_N) )$. Alternatively, one may use the Lorentz transformation, $u(\bp,s)=\Lambda(\bp) u({\bf 0},s)$
 to define the reduced wave function \cite{Hoshizaki68}.

Note that $ \langle JM\ell s\vert JM \lambda_1\lambda_2\rangle $ in
 Eq.~(\ref{eq:JMLL}) 
  is a transformation function between the helicity basis and the orbital-spin 
  basis at fixed $J,M$ \cite{JW}.
%(we here use $\lambda_i$ for the helicity instead of $s_i$ to avoid confusion),
%$\langle ls ms_z\vert JM\rangle $  is the Clebsch-Gordan coefficient for ${\bf J}={\bf L}+{\bf S}$,
Also, 
$ \chi_{ab}(s,s_z)$ in Eq.~(\ref{eq:chi-ab}) is a $2\times 2$ matrix in the spinor space 
with total spin $s=1$ or 0 and its $z$-component $s_z$, and 
$j_\ell(x)$ is a spherical Bessel function. 
Using Eq.~(\ref{eq:orthogonal}), Eq.~(\ref{eq:LS-2})  for large $r$  becomes
\begin{eqnarray}
\psi(\br; \bq,s,s') &=& Z
\sum_{JM} D^J_{M\lambda}(\Omega_{q}) U({\nabla})U(-\nabla) \psi_{JMss'}(\br; q) ,
\quad \lambda = s-s',\\
\label{eq:A3-LST}
 \psi_{JMss'}(\br; q) &\xrightarrow[r>R]{ } &\ \ \phi_{JMss'}^{ [j]}(\br,q)
+\sum_{\tilde{s},\tilde{s}'} \int_0^{\infty} \frac{k^2d k}{2\pi^2}  
\phi_{JM\tilde{s}\tilde{s}'}^{[j]}(\br,k)
\frac{\varepsilon_q+\varepsilon_k}{8 \varepsilon_k^2}
\frac{ \langle \tilde{s} \tilde{s}\vert T^J(k;q)\vert ss'\rangle}
{k^2-q^2 - i\delta}. \nonumber \\
\end{eqnarray}

To evaluate the integral in Eq.~(\ref{eq:A3-LST}), 
we use the following formula \cite{ishizuka,ishizuka2} valid
 for $r>R$ in which  
   $\int_0^{\infty} f(k) {k^{-\ell} j_0(kr)}  k^2 dk= 0$ is satisfied:\footnote{
 Since  the nucleons are non-interacting in the asymptotic region,
 the right hand side of Eq.~(\ref{eq:KG-eq}) is exponentially small for $r>R$.
 This gives a little weaker condition that $\int_0^{\infty} f(k)  j_l(kr)  k^2 dk= 0$,
 which, together with  some properties of the $T$-matrix, leads to the stronger condition used here.}
 \begin{eqnarray}
\label{eq:A3-formula}
\int_0^{\infty}  \frac{k^2d k}{2\pi^2}\frac{ j_{\ell}(kr) f(k)}{k^2-q^2-i\delta} &=&
 i \frac{q}{4\pi} h_{\ell}^{(+)}(qr) f(q).
 \end{eqnarray}
Here $ h_{\ell}^{(\pm)}(x) ( \equiv j_{\ell} (x) \pm i n_{\ell}(x))$
is the spherical Hankel function 
 with $j_0(x) = (\sin x)/x$ and $n_0(x)= - (\cos x)/x$, so that
 $h_{\ell}^{(+)}(qr)$ represents the spherical outgoing-wave. 
   
Then, we obtain
\begin{eqnarray}
 \psi_{JMss'}(\br; q) &\xrightarrow[r > R]{ } & \ \ \phi_{JMss'}^{ [j]}(\br,q)
+i \sum_{\tilde{s},\tilde{s}'} \frac{q}{16\pi \varepsilon_q}
\phi_{JM\tilde{s}\tilde{s}'}^{[h^{(+)}]}(\br,q)
\langle \tilde{s} \tilde{s}\vert T^J(q;q)\vert ss'\rangle \nonumber \\
&=& \sum_{\tilde{s},\tilde{s}'} \left[
\phi_{JM\tilde{s}\tilde{s}'}^{[j]}(\br,q)
A^J_{\tilde{s}\tilde{s}';ss'}(q)
-\phi_{JM\tilde{s}\tilde{s}'}^{[n]}(\br,q)
B^J_{\tilde{s}\tilde{s}';ss'}(q)\right] , \\
A^J(q) &=& 1+ i \frac{q}{16\pi \varepsilon_q} T^J(q;q), \quad
B^J(q)=  \frac{q}{16\pi \varepsilon_q} T^J(q;q) ,
\end{eqnarray}
where $\phi_{JM\tilde{s}\tilde{s}'}^{[n,h^{(+)}]}(\br,q)$ is obtained from 
$ \phi_{JM\tilde{s}\tilde{s}'}^{[j]}(\br,q)$ by 
 the replacement $j_\ell(kr) \rightarrow n_\ell (qr), h_\ell^{(+)} (qr)$.

Using the explicit form of the $T$-matrix given in (\ref{eq:T-form}), we 
 finally obtain
\begin{eqnarray}
X^J_{\ell=J,s} = \hat X_{J s} ,  \ \ \ 
X_{\ell=J\mp 1, s=1}^J  =  O(q) \left(
\begin{array}{cc}
\hat X_{J-1,1} & 0 \\
0 & \hat X_{J+1,1} \\
\end{array}
\right) O^{-1}(q) ,
\end{eqnarray}
with $X$ being either $A$ or $B$, and 
\begin{eqnarray}
\hat A_{\ell s}(q) &=& e^{i\delta_{\ell s}(q)} \cos \delta_{\ell s}(q), \ \ \ 
\hat B_{\ell s}(q) = e^{i\delta_{\ell s}(q)} \sin \delta_{\ell s}(q) ,\\
\frac{\hat A_{\ell s}(q)}{\hat B_{\ell s}(q)} & = & \frac{1}{\tan \delta_{\ell s}(q)}.
\end{eqnarray}

We now have shown that the BS wave function in QCD has an asymptotic form
of the scattering wave of the quantum mechanics at large $r$. To derive this we have only use the unitary of the $S$-matrix below the inelastic threshold, and have identified the phase of the $S$-matrix as the scattering phase shift of the asymptotic BS wave function.   
This observation leads to the important conclusion that the potential defined through the BS wave function, by construction,  gives the correct scattering phase shift at asymptotically large $r$.

%%%%%%%%%%%%%%%%%
\section{Okubo-Marshak Decomposition}
\label{sec:OM} 
In this appendix, we derive the general form of the 
$NN$ potential in the space of two-component spinors, following the argument by 
Okubo and Marshak\cite{okubo}.  The general form of the 2-body potential with derivatives reads
\begin{eqnarray}
V(\br_1, \br_2, \bv_1, \bv_2, \bsigma_1, \bsigma_2, \btau_1, \btau_2 , t),
\end{eqnarray} 
where $\bv_{1,2}=\bp_{1,2}/m_N$. 

There are several conditions to be satisfied by $V$.

\begin{enumerate}
\item Probability conservation:  This leads to the hermiticity of the 
  potential: $V^{\dagger} = V$.
\item Energy-momentum conservation: The energy conservation demands
   that the potential does not depend on time explicitly.  The 
   momentum conservation leads the translational invariance of the potential.
   Thus we have
\begin{eqnarray}
V = V(\br, \bv_1, \bv_2, \bsigma_1, \bsigma_2, \btau_1, \btau_2 ), 
\end{eqnarray}  
 where $\br = \br_1 - \br_2$.
\item  Galilei invariance: The potential is assumed to be 
 independent of the center of mass momentum of the two-body system, which
 leads to
\begin{eqnarray}
V = V(\br, \bv, \bsigma_1, \bsigma_2, \btau_1, \btau_2 ), 
\end{eqnarray}  
where $\bv = \bp/\mu = (\bp_1-\bp_2)/(2\mu)=\bv_1-\bv_2$.
\item Conservation of total-angular momentum:  
The total angular momentum is defined
as ${\bJ}={\bS}+{\bL}$ with
\begin{eqnarray}
  {\bS}=\frac{1}{2}(\bsigma_1+ \bsigma_2), \ \ \ \ 
 {\bL}= \br \times \bp. 
\end{eqnarray}  
The potential is a 
scalar under the spatial rotation.  Then, $V$ is the 
scalar functions of $ \br, \bv, \bsigma_1$ and $\bsigma_2$.
\item Parity invariance: The strong interaction conserves
 parity. Thus $V$ is invariant under reflection, $\br \rightarrow -\br$
  and $\bv \rightarrow - \bv$,
 \begin{eqnarray}
V(\br, \bv, \bsigma_1, \bsigma_2, \btau_1, \btau_2)
= V(-\br, -\bv, \bsigma_1, \bsigma_2, \btau_1, \btau_2).
\end{eqnarray}
\item Time-reversal invariance: The strong interaction
 preserves time-reflection symmetry under $\br \rightarrow \br$,
 $\bv \rightarrow -\bv$, $\bsigma_i \rightarrow -\bsigma_i$, which
 leads to
  \begin{eqnarray}
V(\br, \bv, \bsigma_1, \bsigma_2, \btau_1, \btau_2)
= V(\br, -\bv, -\bsigma_1, -\bsigma_2, \btau_1, \btau_2).
\end{eqnarray}
 \item Fermi statistics: The potential is invariant under
  the permutation of the particle coordinates,
   \begin{eqnarray}
V(\br, \bv, \bsigma_1, \bsigma_2, \btau_1, \btau_2)
 = V(-\br, -\bv, \bsigma_2, \bsigma_1, \btau_2, \btau_1) 
  = V(\br, \bv, \bsigma_2, \bsigma_1, \btau_2, \btau_1),\nonumber\\
\end{eqnarray} 
where parity invariance was used in the second equality.
\item Isospin invariance: The potential is invariant under
 the rotation in isospin space, which leads to two independent
 potentials $V^{I=0,1}$,
\begin{eqnarray}
V = V^{0}(\br, \bv, \bsigma_1, \bsigma_2) P^{\tau}_0
    + V^{1}(\br, \bv, \bsigma_1, \bsigma_2) P^{\tau}_1.
\end{eqnarray}   
\item
Furthermore, $V$ has only the terms  $\sigma_{1}^n \sigma_{2}^m$
 with $(n,m)=(0,0),(1,0),(0,1),(1,1)$. The other higher order
  terms can be always reduced to the above form because of the 
  property of the Pauli matrices: 
  $\sigma_{i} \sigma_{j}= \delta_{ij} + i \varepsilon_{ijk} \sigma_k$. 
 \end{enumerate}
 
 Then, the terms which have Pauli matrices and satisfy the 
 above constraints  are restricted only to the 
 following combinations:
\begin{eqnarray}
& & \bsigma_1 \cdot \bsigma_2, \quad
(\bsigma_1 + \bsigma_2 ) \cdot {\bL} ,\quad
(\bsigma_1 \cdot \br )(\bsigma_2 \cdot \br ) ,\quad \\
& & (\bsigma_1 \cdot \bv )(\bsigma_2 \cdot \bv ) ,\quad
(\bsigma_1 \cdot {\bL} )(\bsigma_2 \cdot {\bL} ) .
\end{eqnarray}  
 It is sometimes convenient to reorganize the above 5 terms into
 the following hermitian operators: 
\begin{eqnarray}
\label{sec:B-5op}
& &\bsigma_1 \cdot \bsigma_2, \quad 
S_{12} \equiv 3(\bsigma_1 \cdot \hat{\br} )(\bsigma_2 \cdot \hat{\br} ) -
\bsigma_1 \cdot \bsigma_2 ,\quad \nonumber \\
& & {\bL} \cdot {\bS} , \nonumber \\
& & P_{12}\equiv (\bsigma_1 \cdot \bv )(\bsigma_2 \cdot \bv ) ,\quad
W_{12}\equiv Q_{12} -\frac{1}{3} (\bsigma_1\cdot \bsigma_2) {\bL}^2,
\end{eqnarray} 
where
\begin{eqnarray}
Q_{12}&\equiv& \frac{1}{2} \left[(\bsigma_1 \cdot {\bL} )(\bsigma_2 \cdot {\bL} ) 
+(\bsigma_2 \cdot {\bL} )(\bsigma_1 \cdot {\bL} ) \right] .
\end{eqnarray}
In $Q_{12}$ the spins need to be symmetrized to make
 it hermitian since $L_i$ and $L_j$ do not commute with each other.
Note that the term such as $({\bS}\cdot {\bL})^2$ can be decomposed into
 $Q_{12}$, ${\bS}\cdot {\bL}$ and spin-independent ${\bL}^2$ term.

We decompose $V^0$ and $V^{1}$ in terms of the above
 operators with coefficients $V^I_A$  
 ($I= (0, 1)$, $A=(0, \sigma, T, LS, P, W$))
  which are the 
  scalar function made of $\br$ and $\bv$ satisfying the 
  general constraints;
 \begin{eqnarray}
V^I_A= V^I_A (\br^2, \bv^2, {\bL}^2).
\end{eqnarray}  
Note that the scalar $(\br \cdot \bp)^2$ can be written 
by $\br^2 \bp^2$ and ${\bL}^2$.

Combining all, we arrive at the general decomposition
 given in \S\ref{sec:OM-potential}.
  
%%%%%%%%%%%%
\section{Matrix Element of the Potential}  
\label{sec:matrix_element}
In this appendix, we consider the partial wave decomposition of the general form of the $NN$ potential.
At given $J$, there are 2 distinct states, the spin-singlet ($s=0$)
 state and the spin-triplet ($s=1$) state.
We now consider how the five operators in Eq.~(\ref{sec:B-5op}) act on these states.

The singlet  state is denoted as $^1J_J$, %by the notation that $^{2S+1} L_J$,
since it has $s=0$ and $J=\ell$. 
 The fact that $I+\ell+s$ must be odd to satisfy fermion anti-symmetry gives $I=0$ for odd $J$ and $I=1$ for even $J$. The eigenstate with $J_z=M$ can be easily obtained as
\begin{eqnarray}
\vert ^1J_J, M\rangle &=& \vert M, 0\rangle_{J,0},
\end{eqnarray}
where we use the short-handed notation,
$\vert J_z, s_z\rangle_{J,s} = \vert J, J_z\rangle \otimes \vert s, s_z\rangle $.

The spin-triplet state is classified into 3 types:  $^3J_J$, $^3(J\pm 1)_J$. For the first one, $I=0$ (even $J$) or $I=1$ (odd $J$), and vice versa for other two types. 
 By the Wigner-Eckart theorem,
the matrix elements of the five operators do not depend on $J_z$. Therefore it is enough to know eigenstates with $J_z=J$ only. Explicitly we have
\begin{eqnarray}
\vert ^3J_J, J\rangle &=& \frac{1}{\sqrt{J+1}}\left\{ \vert J-1,1\rangle_{J,1} 
- \sqrt{J}\vert J,0\rangle_{J,1}\right\} \ \ ,\\
\vert ^3(J-1)_J, J\rangle &=&  \vert J-1,1\rangle_{J-1,1} \ \ , \\
\vert ^3(J+1)_J, J\rangle &=& \frac{1}{\sqrt{(J+1)(2J+3)}} \Bigl\{ \vert J-1,1\rangle_{J+1,1} 
\nonumber \\
&+&\sqrt{2J+1}\left[ \sqrt{(J+1)}\vert J+1,-1\rangle_{J+1,1}-\vert J,0\rangle_{J+1,1}\right]\Bigr\}.
\end{eqnarray}

Using these eigenstates, it is easy to see
\begin{eqnarray}
\bsigma_1\cdot\bsigma_2 &=& 2s(s+1)-3 = -3,\ 1,\ 1,\ 1 \ \ \ \ ,\\
{\bL}\cdot {\bS} &=& \frac{J(J+1)-\ell(\ell+1)-s(s+1)}{2} = 0,\ -1,\  J-1,\ -(J+2) \ \ \ \ , \\
W_{12} &=& 0,\quad  -\frac{(2J-1)(2J+3)}{3}, \quad\frac{(J-1)(2J-3)}{3},\quad \frac{(J+2)(2J+5)}{3} ,
\end{eqnarray}
for $^1J_J$, $^3J_J$, $^3(J-1)_J$ and $^3(J+1)_J$, respectively.

For $S_{12}$ and $P_{12}$ results are more complicated due to the mixing between $^3(J-1)_J$ and $^3(J+1)_J$. After a little algebra we obtain
\begin{eqnarray}
S_{12} &=& 0, \ 2, \ \left( \begin{array}{cc}
 -\dfrac{2(J-1)}{2J+1}, & \dfrac{6\sqrt{J(J+1)}}{2J+1}  \\
 \\
 \dfrac{6\sqrt{J(J+1)}}{2J+1}, & -\dfrac{2(J+2)}{2J+1} \\
\end{array}\right)  , \\
\mu^2 P_{12} &=& 0, \ 2p_J^2, \ \left( \begin{array}{cc}
 -\dfrac{2(J-1)}{2J+1}p_{J-1}^2, & \dfrac{6\sqrt{J(J+1)}}{2J+1} p_+^2\\
 \\
 \dfrac{6\sqrt{J(J+1)}}{2J+1}p_-^2 ,
 & -\dfrac{2(J+2)}{2J+1}p_{J+1}^2 \\
\end{array}\right) ,
\end{eqnarray}
where 
\begin{eqnarray}
p_{\ell}^2 &=& p_r^2 -i\frac{2}{r} p_r + \frac{\ell (\ell +1)}{r^2} \equiv \mu^2 v_{\ell}^2, \\
p_+^2 &=& \left(p_r -i\dfrac{J+1}{r}\right)\left(p_r -i\dfrac{J+2}{r}\right) \equiv \mu^2 v_+^2, \\
p_-^2 &=& \left(p_r +i\dfrac{J}{r}\right)\left(p_r +i\dfrac{J-1}{r}\right) \equiv \mu^2 v_-^2,
\end{eqnarray}
with $\ell = J \pm 1$ and $i p_r = \partial/(\partial r) $.

Using these results, we obtain the potential for each channel:  We have
\begin{eqnarray}
V[^1J_J] &=& V_0^I(r^2, v_J^2, \hat J^2) + V_\sigma^I(r^2, v_J^2, \hat J^2), \qquad
\hat J^2 = J(J+1) 
\end{eqnarray}
for the $^1J_J$ state, and 
\begin{eqnarray}
V[^3J_J] &=& V_0^{\bar I}(r^2, v_J^2, \hat J^2) -3 V_\sigma^{\bar I}(r^2, v_J^2, \hat J^2)
-V_{LS}^{\bar I}(r^2, v_J^2, \hat J^2)+2 V_{T}^{\bar I}(r^2, v_J^2, \hat J^2)\nonumber \\
&-&\frac{(2J-1)(2J+3)}{3}V_{W}^{\bar I}(r^2, v_J^2, \hat J^2)+\{ V_{P}^{\bar I} (r^2, v_J^2, \hat J^2), \ v_J^2 \}
\end{eqnarray}
for the $^3J_J$ state, where $\bar I = 1 - I$.

For $^3(J\mp 1)_J$, the result is more involved:
\begin{eqnarray}
V[^3(J\mp 1)_J] &=& \left(\begin{array}{cc}
V_{--} & V_{-+} \\
V_{+-} & V_{++} \\
\end{array}\right) ,
\end{eqnarray}
where
\begin{eqnarray}
V_{--} &=& V_0^I (r^2, v_{J_-}^2, \hat J_-^2) - 3V_\sigma^I (r^2, v_{J_-}^2, \hat J_-^2)
+(J-1)V_{LS}^I (r^2, v_{J_-}^2, \hat J_-^2) \nonumber \\
 & & -\frac{(J-1)(2J-3)}{3}V_W^I (r^2, v_{J_-}^2, \hat J_-^2)\nonumber\\
&& -\frac{J-1}{2J+1}\left[ 2V_{T}^I (r^2, v_{J_-}^2, \hat J_-^2)+ \{V_{P}^I (r^2, v_{J_-}^2, \hat J_-^2), v_{J_-}^2\} \right] , \\
V_{++} &=& V_0^I (r^2, v_{J_+}^2, \hat J_+^2) - 3V_\sigma^I (r^2, v_{J_+}^2, \hat J_+^2)
-(J+2)V_{LS}^I (r^2, v_{J_+}^2, \hat J_+^2) \nonumber \\
& &  +\frac{(J+2)(2J+5)}{3}V_W^I (r^2, v_{J_+}^2, \hat J_+^2)\nonumber\\
&&-\frac{J+2}{2J+1}\left[ 2V_{T}^I (r^2, v_{J_+}^2, \hat J_+^2)+ \{V_{P}^I (r^2, v_{J_+}^2, \hat J_+^2),
v_{J_+}^2\} \right] ,\\
V_{-+} &=& \frac{3\sqrt{J(J+1)}}{2(2J+1)}\left[
2V_{T}^I (r^2, v_{J_+}^2, \hat J_+^2)+ 2V_{T}^I (r^2, v_{J_-}^2, \hat J_-^2) \right. \nonumber \\
& &\ \ \ \ \  \ \ \ \  \ \ \ \ \ \ \ \  \ \ \ \ \ \ \ \ 
\left. + v_+^2 V_{P}^I (r^2, v_{J_-}^2, \hat J_-^2) + v_+^2 V_{P}^I (r^2, v_{J_+}^2, \hat J_+^2)  v_+^2
\right] ,\\
V_{+-} &=& \frac{3\sqrt{J(J+1)}}{2(2J+1)}\left[
2V_{T}^I (r^2, v_{J_+}^2, \hat J_+^2)+ 2V_{T}^I (r^2, v_{J_-}^2, \hat J_-^2) \right. \nonumber \\
& & \ \ \ \ \  \ \ \ \  \ \ \ \ \ \ \ \  \ \ \ \ \ \ \ \ 
 \left. +  v_-^2V_{P}^I (r^2, v_{J_-}^2, \hat J_-^2) + v_-^2  V_{P}^I (r^2, v_{J_+}^2, \hat J_+^2)
\right] ,
\end{eqnarray}
where $J_\pm = J\pm 1$ and $\hat J_\pm^2 = J_\pm(J_\pm+1)$.
Note that $V_{+-} = (V_{-+})^\dagger$ with $r^2$ from the integration measure.

\section{Heat-Kernel Representation of the Green's Function}
\label{sec:heat-kernel}

We define the heat kernel ${\cal K}(t, \bx)$ through the initial value
problem as
\begin{equation}
  \frac{\partial}{\partial t}
  {\cal K}(t, \bx)
  =
  \nabla^2
  {\cal K}(t, \bx),
  \hspace{2em}
  \lim_{t\to 0^+}
  {\cal K}(t, \bx)
  =
  \delta_{\rm lat}(\bx),
\end{equation}
where  $ \delta_{\rm  lat}(\bx)  \equiv  \frac1{L^3}
\sum_{\bns \in \boldsymbol{Z}^3} e^{2\pi i \bns \cdot \bxs /L}$ denotes
the delta function in a periodic box of spatial extension $L$.
${\cal K}(t,\bx)$ is explicitly expressed as
\begin{equation}
  {\cal K}(t,\bx)
  =
  e^{t \nabla^2}
  \delta_{\rm lat}(\bx)
  =
  \frac1{L^3}
  \sum_{\bns \in \boldsymbol{Z}^3}
  \exp\left(
  - t (2\pi/L)^2 \bn^2 + 2\pi i \bn \cdot \bx/L
  \right).
  \label{eq.heat.kernel.explicit}
\end{equation}
For convenience, we define a modified heat kernel 
$\tilde {\cal K}(t, \bx; k^2)$ as
\begin{equation}
  \tilde {\cal K}(t, \bx; k^2)
  \equiv
  e^{t k^2}
  {\cal K}(t, \bx),
\end{equation}
which is the solution to the equation that
\begin{equation}
  \frac{\partial}{\partial t}
  \tilde {\cal K}(t, \bx; k^2)
  =
  \left(
  \nabla^2 + k^2
  \right)
  \tilde {\cal K}(t, \bx; k^2),
  \hspace{2em}
  \lim_{t\to 0^+}
  \tilde {\cal K}(t, \bx; k^2)
  =
  \delta_{\rm lat}(\bx).
\end{equation}
An integration from $0$ to $s$ leads us to
\begin{equation}
  \tilde {\cal K}(s, \bx; k^2) = \delta_{\rm lat}(\bx)
  +
  \int_0^s dt
  \left( \nabla^2 + k^2 \right)
  \tilde {\cal K}(t, \bx; k^2).
\end{equation}
%which is arranged as
%\begin{equation}
%  - \tilde K(s, \bx; k^2)
%  +
%  \left( \nabla^2 + k^2 \right)
%  \int_0^s dt\;
%  \tilde K(t, \bx; k^2)
%  =
%  -\delta_{\rm lat}(\bx).
%\end{equation}
This identity gives an expression for Green's function as
\begin{eqnarray}
  G(\bx; k^2)
  &\equiv&
  -\left( \nabla^2 + k^2 \right)^{-1}
  \tilde {\cal K}(s, \bx; k^2)
  +
  \int_0^s dt\;
  \tilde {\cal K}(t, \bx; k^2)
  \label{eq.green.function.heat.kernel}
  \nonumber\\
  &=&
  -
  e^{s k^2}
  \left( \nabla^2 + k^2 \right)^{-1}
  {\cal K}(s, \bx)
  +
  \int_0^s dt\;
  e^{t k^2}
  {\cal K}(t, \bx).
\end{eqnarray}
By inserting Eq.~(\ref{eq.heat.kernel.explicit}), we have
\begin{eqnarray}
\hspace*{-7mm}      G(\bx; k^2)
&=&  \frac{e^{s k^2}}{L^3}
  \sum_{\bns \in \boldsymbol{Z}^3}
  \frac{
    \exp\left(
    - s (2\pi/L)^2 \bn^2
    + 2\pi i \bn\cdot\bx/L
    \right)
  }
  {
    (2\pi/L)^2 \bn^2 - k^2
  }\nonumber\\
   && +
    \int_0^s dt\;
    \frac{e^{t k^2}}{(4\pi t)^{3/2}}
    \sum_{\bps \in \boldsymbol{Z}^3}
    \exp\left(
    -\frac1{4t}
    \left( \bx - \bp L \right)^2
    \right),   \label{eq.convergent.green}
\end{eqnarray}
where, in the second term, we used Poisson's summation formula
\begin{equation}
  \sum_{n \in \boldsymbol{Z}}
  \exp\left(
  -\frac{n^2}{2\beta} + in\theta
  \right)
  =
  \sqrt{2\pi\beta}
  \sum_{p \in \boldsymbol{Z}}
  \exp\left(
  -\frac{\beta}{2}
  \left(\theta + 2\pi p\right)^2
  \right).
\end{equation}
The   convergence   of  the   summations   and   the  integration   in
Eq.~(\ref{eq.convergent.green}) is quite good
%
%except that, for $\vec p = \vec 0$,
%
except at $\bp = {\bf 0}$, where
the integration in the second term  has to be done analytically for
 $t \sim 0$.  For this purpose, we use the formula
\begin{equation}
  \frac1{4\pi r}
  =
  \int_0^{\infty}
  \frac{dt}{(4\pi t)^{3/2}}
  \exp\left( - \frac{r^2}{4t} \right),
\end{equation}
%The intergration can be arranged as
%
and the integration can be evaluated as
\begin{eqnarray}
  \lefteqn{
    \int_0^{s}
    \frac{dt\; e^{t k^2}}{(4\pi t)^{3/2}}
    \exp\left(-\frac1{4t} \bx^2 \right)
  }
  \nonumber\\
  &=&
  \frac1{4\pi |\bx|}
  -
  \int_0^{s}
  \frac{dt}{(4\pi t)^{3/2}}
  \left( e^{t k^2} - 1 \right) \exp\left( -\frac{\bx^2}{4t}\right)
  -
  \int_s^{\infty}
  \frac{dt}{(4\pi t)^{3/2}}
  \exp\left( - \frac{\bx^2}{4t} \right).\nonumber \\
\end{eqnarray}
Note that  $s$ dependences in Eq.~(\ref{eq.green.function.heat.kernel})
and Eq.~(\ref{eq.convergent.green}) cancel out  on the right-hand side,
and  $s$  plays a  role  of  the cutoff  $\lambda$  of Eq.~(D.2)  in
Ref.~\citen{luescher}  as  $s   \sim  1/\lambda^2$.   It  controls  the
convergence    of   the    summation    in   the    first   term    in
Eq.~(\ref{eq.convergent.green}).


\begin{thebibliography}{99}
 
\bibitem{Nijmegen_data}
\texttt{http://nn-online.org/}

\bibitem{NN-review}
M. Taketani et al., Prog. Theor. Phys. Suppl. No.~39 (1967), 1.\\
N. Hoshizaki et al., Prog. Theor. Phys. Suppl. No.~42 (1968), 1. \\ 
G. E. Brown and A. D. Jackson, {\it Nucleon-nucleon Interaction}
(North-Holland, Amsterdam, 1976).\\
R.~Machleidt, Adv.\ Nucl.\ Phys.\ {\bf 19} (1989), 189.\\
R. Machleidt and I.~Slaus, J. of  Phys. G {\bf 27} (2001), R69.

\bibitem{CD-Bonn}
R.~Machleidt, Phys.\ Rev.\ C {\bf 63} (2001), 024001.

\bibitem{Wiringa:1994wb}
R.~B.~Wiringa, V.~G.~J.~Stoks and R.~Schiavilla, Phys.\ Rev.\ C {\bf 51} (1995), 38.
 
\bibitem{Stoks:1994wp}
V.~G.~J.~Stoks, R.~A.~M.~Klomp, C.~P.~F.~Terheggen and J.~J.~de Swart,
Phys.\ Rev.\ C {\bf 49} (1994), 2950.

\bibitem{ChPT-1}
S.~Weinberg, Phys.\ Lett.\ B {\bf 251} (1990), 288;
Nucl.\ Phys.\ B {\bf 363} (1991), 3.\\
P.~F.~Bedaque and U.~van Kolck, Ann.\ Rev.\ Nucl.\ Part.\ Sci.\ {\bf 52} (2002), 339.

\bibitem{ChPT-2}
Reviewed in  R.~Machleidt, arXiv:0704.0807.\\
E.~Epelbaum, H.~W.~Hammer and U.~G.~Meissner, arXiv:0811.1338.

\bibitem{yukawa}
H.~Yukawa, Proc. Math.-Phys. Soc. Jpn. {\bf 17} (1935), 48.

\bibitem{jastrow}
R. Jastrow, Phys. Rev. {\bf 81} (1951), 165.

\bibitem{VJ}
R.~Tamagaki et al., Prog.\ Theor.\ Phys.\ Suppl.\ No.~112 (1993), 1.\\
H. Heiselberg and V. Pandharipande, Annu. Rev. Nucl. Part. Sci. {\bf 50} (2000), 481.\\
J. M. Lattimer and M. Prakash, Phys. Rep. {\bf 333} (2000), 121.
 
\bibitem{QQ_review}
F.~Myhrer and J. Wroldsen, Rev. Mod. Phys. {\bf 60} (1988), 629.\\
M.~Oka, K.~Shimizu and K.~Yazaki, Prog. Theor. Phys. Suppl. No.~137 (2000), 1.\\
Y.~Fujiwara, Y.~Suzuki and C.~Nakamoto, Prog.\ Part.\ Nucl.\ Phys.\ {\bf 58} (2007), 439.

\bibitem{nambu57}
Y. Nambu, Phys. Rev. {\bf 106} (1957), 1366. 

\bibitem{skyrme}
A.~Jackson, A.~D.~Jackson and V.~Pasquier, Nucl.\ Phys.\ A {\bf 432} (1985), 567.\\
H.~Yabu and K.~Ando, Prog.\ Theor.\ Phys.\ {\bf 74} (1985), 750.

\bibitem{core_quark}
S. Otsuki, R. Tamagaki and M. Yasuno, Prog. Theor. Phys. Suppl. Extra Number (1965), 578.\\
S. Machida and M. Namiki, Prog. Theor. Phys. {\bf 33} (1965), 125.\\
V.~G.~Neudachin, Yu.~F.~Smirnov and R.~Tamagaki, Prog.\ Theor.\ Phys.\ {\bf 58} (1977), 1072.\\
D.~A.~Liberman, Phys.\ Rev.\ D {\bf 16} (1977), 1542.\\
C.~DeTar, Phys.\ Rev.\ D {\bf 19} (1979), 1451.\\
M.~Oka and K.~Yazaki, Phys.\ Lett.\ B {\bf 90} (1980), 41;
Prog.\ Theor.\ Phys.\ {\bf 66} (1981), 556; Prog.\ Theor.\ Phys.\ { \bf 66} (1981), 572.\\
H.~Toki, Z.\ Phys.\ A {\bf 294} (1980), 173.\\
A.~Faessler, F.~Fernandez, G.~Lubeck and K.~Shimizu, Phys.\ Lett.\ B {\bf 112} (1982), 201.

\bibitem{luescher}
M.~L\"{u}scher, Nucl.\ Phys.\ B {\bf 354} (1991), 531.

\bibitem{suganuma}
D.~Arndt, S.~R.~Beane and M.~J.~Savage, Nucl.\ Phys.\ A {\bf 726} (2003), 339.\\
T.~T.~Takahashi, T.~Doi and H.~Suganuma, 
AIP Conf.\ Proc.\ {\bf 842} (2006), 249, hep-lat/0601006.

\bibitem{forcrand}
P.~de~Forcrand and M.~Fromm, arXiv:0907.1915.

\bibitem{Fukugita:1994ve}
M.~Fukugita, Y.~Kuramashi, M.~Okawa, H.~Mino and A.~Ukawa,
Phys.\ Rev.\ D {\bf 52} (1995), 3003, hep-lat/9501024.

\bibitem{NPLQCD}
S.~R.~Beane, P.~F.~Bedaque, K.~Orginos and M.~J.~Savage,
Phys.\ Rev.\ Lett.\ {\bf 97} (2006), 012001, hep-lat/0602010. 

\bibitem{Ishii:2006ec}
N.~Ishii, S.~Aoki and T.~Hatsuda,
Phys.\ Rev.\ Lett.\ {\bf 99} (2007), 022001, nucl-th/0611096.

\bibitem{Aoki:2008hh}
S.~Aoki, T.~Hatsuda and N.~Ishii,
Comput. Sci. Disc. {\bf 1} (2008), 015009, arXiv:0805.2462.

\bibitem{BEC-BCS}
 T. Koehler, K. Goral and P. S. Julienne,
 Rev. Mod. Phys. {\bf 78} (2006), 1311, cond-mat/0601420.

\bibitem{Kuramashi:1995sc}
Y.~Kuramashi, Prog.\ Theor.\ Phys.\ Suppl.\ No.~122 (1996), 153, hep-lat/9510025.

 \bibitem{Nemura:2008sp}
H.~Nemura, N.~Ishii, S.~Aoki and T.~Hatsuda,
Phys.\ Lett.\ B {\bf 673} (2009), 136, arXiv:0806.1094.

\bibitem{okubo}
S. Okubo and R.~E. Marshak, Ann. of Phys. {\bf 4} (1958), 166.
 
\bibitem{TW67}
R. Tamagaki and W. Watari, Prog. Theor. Phys. Suppl. No.~39 (1967), 23.

 \bibitem{Fujita:1957zz}
J.~Fujita and H.~Miyazawa, Prog.\ Theor.\ Phys.\ {\bf 17} (1957), 360.

\bibitem{Weinberg:1992yk}
S.~Weinberg, Phys.\ Lett.\ B {\bf 295} (1992), 114, hep-ph/9209257. 

\bibitem{Pieper:2001ap}
S.~C.~Pieper, V.~R.~Pandharipande, R.~B.~Wiringa and J.~Carlson,
Phys.\ Rev.\ C {\bf 64} (2001), 014001, nucl-th/0102004.

\bibitem{Akmal:1998cf}
A.~Akmal, V.~R.~Pandharipande and D.~G.~Ravenhall,
 Phys.\ Rev.\ C {\bf 58} (1998), 1804, nucl-th/9804027.

\bibitem{Furumoto:2009zz}
T.~Furumoto, Y.~Sakuragi and Y.~Yamamoto,
Phys.\ Rev.\ C {\bf 79} (2009), 011601.

\bibitem{luescher-CMP}
M.~L\"{u}scher, Commun. Math. Phys. {\bf 105} (1986), 153.

\bibitem{NZ}
K. Nishijima, Phys. Rev. {\bf 111} (1958), 995.\\
W. Zimmermann, Nuovo Cim. {\bf 10} (1958), 597.\\
R. Haag, Phys. Rev. {\bf 112} (1958), 669.\\
For a brief review, see W. Zimmermann, MPI-PAE/PTh-61/87 (1987), unpublished.

\bibitem{Labrenz:1996jy}
J.~N.~Labrenz and S.~R.~Sharpe, Phys.\ Rev.\ D {\bf 54} (1996), 4595, hep-lat/9605034.
 
\bibitem{savage2}
S.~R.~Beane and M.~J.~Savage, Phys. Lett. B {\bf 535} (2002), 177.

\bibitem{Hatsuda:1989bi}
T.~Hatsuda, Nucl.\ Phys.\ B {\bf 329} (1990), 376.

\bibitem{ishizuka}
S.~Aoki et al. (CP-PACS Collaboration), Phys. Rev. D {\bf 71} (2005), 094504.
 
\bibitem{ishizuka2}
N.~Ishizuka, arXiv:0910.2772.

\bibitem{Ishii:2009zr}
N.~Ishii, S.~Aoki and T.~Hatsuda (PACS-CS Collaboration), PoS(LATTICE 2008)155, arXiv:0903.5497.

\bibitem{kuramashi2}
M.~Fukugita, Y.~Kuramashi, M.~Okawa, H.~Mino and A.~Ukawa,
Phys. Rev. D {\bf 52} (1995), 3003.

\bibitem{Epelbaum:2005pn}
E.~Epelbaum, Prog.\ Part.\ Nucl.\ Phys.\ {\bf 57} (2006), 654,
nucl-th/0509032.
 
\bibitem{murano}
S.~Aoki, J.~Balog, T.~Hatsuda, N.~Ishii, K.~Murano, H.~Nemura and P.~Weisz,
PoS(LATTICE 2008)152, arXiv:0812.0673.

\bibitem{ope}
S.~Aoki, J.~Balog and P.~Weisz, Prog. Theor. Phys. {\bf 121} (2009), 1003.

\bibitem{Nemura:2009kc}
H.~Nemura, N.~Ishii, S.~Aoki and T.~Hatsuda (PACS-CS Collaboration),
PoS(LATTICE 2008)156, arXiv:0902.1251.
 
\bibitem{Kuramashi:2008tb}
Y.~Kuramashi, PoS(LATTICE 2008)018, arXiv:0811.2630.

\bibitem{JW}
M.~Jacob and G.~C.~Wick, Ann. of Phys. {\bf 7} (1959), 404.

\bibitem{Lin:2001ek}
C.~J.~D.~Lin, G.~Martinelli, C.~T.~Sachrajda and M.~Testa,
Nucl.\ Phys.\ B {\bf 619} (2001), 467.

\bibitem{Hoshizaki68}
N. Hoshizaki, Prog. Theor. Phys. Suppl. No.~42 (1968), 107. \\ 

\end{thebibliography}
\end{document}